\documentclass[12pt,a4paper,final]{iopart}

\usepackage{iopams}
\usepackage{graphicx}
\usepackage{subfigure}
\usepackage{graphicx,subfigure,color}
\usepackage[breaklinks=true,colorlinks=true,linkcolor=blue,urlcolor=blue,citecolor=blue]{hyperref}
\usepackage{cite}
\usepackage{soul}

	\newcommand{\ket}[1]{\left| #1 \right\rangle}
	\newcommand{\bra}[1]{\left\langle #1 \right|}

\begin{document}

\title[Quantum repeater using three-level atomic...]{Quantum repeater using three-level atomic states in the presence of dissipation: stability of entanglement}

\author{M Ghasemi$^1$}
\ead{m.ghasemi@stu.yazd.ac.ir}

\author{M K Tavassoly$^1$}
\address{$^1$Atomic and Molecular Group, Faculty of Physics, Yazd University, Yazd 89195-741, Iran}
\ead{mktavassoly@yazd.ac.ir}

\begin{abstract}
 In this paper we want to investigate the possibility of transferring entanglement to two three-level separable atomic states over large distance using the quantum repeater protocol. In detail, our model consists of eight $\Lambda$-type three-level atoms where only the pairs (1,2), (3,4), (5,6) and (7,8) are prepared in maximally entangled states. Performing suitable interaction between non-entangled three-level atoms (2,3) and (6,7) in two-mode cavities with photon leakage rates $\kappa$, $\kappa'$ in the presence of spontaneous emission leads to producing entanglement between atoms (1,4) and (5,8), separately. Finally, the entanglement between atoms (1,8) is successfully produced by performing interaction between atoms (4,5) while spontaneous emission is considered in a dissipative cavity. In the continuation, the effects of detuning, dissipation and initial interaction time are considered on negativity and success probability of the processes. The maxima of negativity are decreased by increasing the detuning, in most cases. Also, the time evolution of negativity is non-periodic in the presence of dissipation. Increasing the initial interaction time has a constructive effect on negativity in all considered cases. The oscillations of negativity are destroyed as time goes on and the produced entanglement is stabled. The success probability of entangled state of atoms (1,8) is tunable by controlling the detuning and dissipation. We show that via justifying the involved parameters one can arrive at conditions in which the decoherence effects are fully disappeared; as a result an ideal quantum repeater can be achieved while atomic and field dissipations are taken into account.
\end{abstract}

\vspace{2pc}

\submitto{J. Phys. B}

\maketitle

\section{Introduction}
Quantum entanglement plays a fundamental role in different branches of quantum information science and technologies, in particular in quantum communication
 \cite{Pan2001,Han2006,Jen2012}, quantum cryptography \cite{Jennewein2000,Wang2005} and quantum key distribution \cite{Curty2004}, especially for long distances.
The Jaynes-Cummings \cite{Jaynes1963} model (the interaction between a two-level atom and a single-mode quantized field) is a well-known approach for entanglement production.
In addition, this model is expandable to multi-level multi-atom (Tavis-Cummings model \cite{Tavis1968}) interacting with multi-mode field. Tavis-Cummings model has been frequently considered
 in the literature  \cite{Faghihi2013,Bashkirov2010,Villas2003,Baghshahi1,Baghshahi2,Baghshahi3,Baghshahi4} due to importance of multi-level systems.\\
The entangled states can be transferred over long distances by the quantum repeater protocol which at first has been proposed in \cite{Briegel1998}. Following this pioneering work, a quantum repeater protocol based on photonic cluster-state machine guns has been reported in \cite{Azuma2015}. The phase noise in long optical fibers that is very useful for quantum repeater applications is investigated in \cite{Minavr2008}. The mechanism of an efficient quantum repeater, employing double-photon guns, for long-distance optical quantum communication has been demonstrated in \cite{Kok2003}. Recently, by considering an array of photons $(1,2,\cdots,8)$  the authors in \cite{Xu2017} have reported an experiment  which demonstrates swapping the entanglement to the distant separable photons (1,8) via performing Bell-state measurement (BSM) method on pairs (2,3), (4,5) and (6,7) with the help of quantum repeater concept. Also, purification for a linear optical quantum repeater experimentally has been considered in Ref. \cite{Chen2017}.
An experimental realization of entanglement concentration using two polarization entangled photon pairs has been reported through which a quantum repeater is demonstrated \cite{Zhao2003}.
Also, the quantum repeater node has been considered, experimentally in \cite{Yuan2008}.\\
In quantum repeater, the long  distance between a separable two-partite quantum state is divided into short (entangled) parts, whereas these entangled parts
 are separable from each other. Basically, using the entanglement swapping process \cite{Xue1,Wu2013,Yonac2006,Man2006,Yang2,Ghasemi3,Ghasemi4},
 the entanglement can be produced between the separable parts. To do this task, besides the BSM method, performing the interaction between non-entangled segments quantum electrodynamics (QED) method is frequently used for swapping the entanglement \cite{Pakniat1,Zhang2}. Also, quantum memories allow to store swapped entanglement or purified it in any segment \cite{Tittel2010,Gouet2012}. A quantum repeater protocol has been recently proposed using the qubit atomic system \cite{Yi2017}. Moreover,
analysis shows that qutrit schemes for quantum key distribution are more robust than the qubit schemes \cite{Dun2003}. Also,
quantum cryptography with three-level systems has better security compared with two-level systems \cite{Bechmann2000}.
A higher degree of security obtainable in entanglement based on multi-level quantum cryptography has been studied in \cite{Bourennane2001}.
The authors in Ref. \cite{Kaszlikowski2000} have reported that \textit{"violations of local realism are stronger for two maximally entangled quNits $(3\leq N\leq9)$
than for two qubits and that they increase with N"}. \\
Considering the above-mentioned facts, we motivate to introduce a quantum repeater protocol with two distant three-level atoms, while in the entanglement transfer processes the losses effects are also taken into account.
In detail, we consider eight $\Lambda$-type three-level atoms where the pairs (1,2), (3,4), (5,6) and (7,8) are prepared in maximally entangled states.
By performing the interaction between dissipative atoms (2,3) and (6,7) in the presence of photon leakage from cavities, the entangled atoms (1,4) and (5,8) are produced. Finally, the separable atoms (1,8) are entangled by performing interaction between atoms (4,5).
Due to the unavoidable fact  that, there exists entanglement attenuation which originates from the interaction of any system with its environment, to close such interaction models to reality, the dissipation effects such as spontaneous emission and photon leakage from non-perfect cavities have been considered in our presentation \cite{yang3,Di2008}.
The negativity measure of entanglement \cite{Akhtarshenas2007} and success probability are deduced and the effects of detuning, dissipation and initial interaction time are discussed in detail.\\
 This paper is organized as follows: We explain our model and present our calculations in Sec. 2. The numerical results of negativity and success probability are analysed in Sec. 3. Finally, a summary and conclusions are collected in Sec. 4.
\section{Quantum repeater protocol}\label{model}

\begin{figure}
 \centering
 \subfigure[\label{fig.Fig1a} \ Model]{\includegraphics[width=0.35\textwidth]{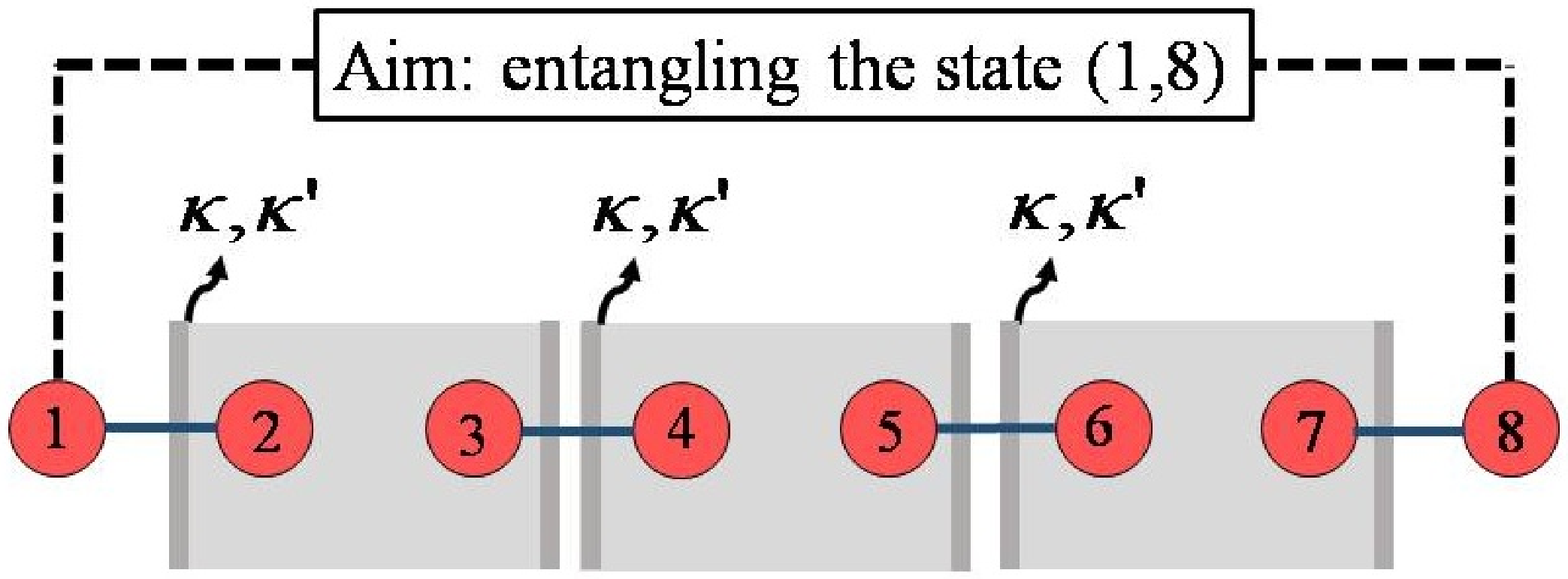}}
 \hspace{0.05\textwidth}
  \subfigure[\label{fig.Fig1b} \ The diagram of three-level $\Lambda$-type atoms]{\includegraphics[width=0.50\textwidth]{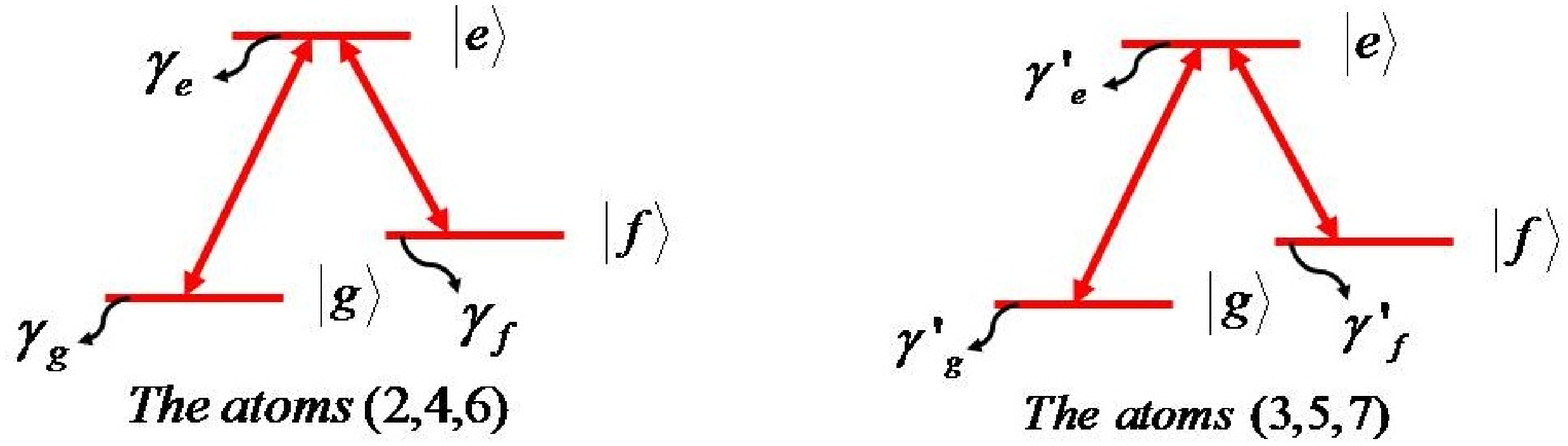}}
 \caption{\label{fig.model} (a) The entanglement is produced between two atoms (1,8) by using the quantum repeater protocol in the presence of photon leakage rates $\kappa$ and $\kappa'$ associated to the two-mode fields from non-perfect cavities. (b) The three-level $\Lambda$-type atoms with spontaneous emission.}
  \end{figure}
 In our model, we have eight $\Lambda$-type three-level atoms where the pairs (1,2), (3,4), (5,6) and (7,8) are prepared in atomic Bell states shown in figure \ref{fig.Fig1a}. The purpose of this paper is producing entanglement between non-entangled distant atoms (1,8). The separable pairs (1,4) and (5,8) can be respectively entangled by performing interaction between atoms (2,3) and (6,7) (see figure \ref{fig.Fig1b}) in the two-mode non-perfect cavities while spontaneous emission is also considered. Finally, the state of atoms (1,8) is converted to entangled state by performing interaction between atoms (4,5).\\
 We suppose that the atoms (1,2) and (3,4) are in the initial state $\ket{\Psi}_{1,2}\otimes\ket{\Psi}_{3,4}$, where
   \begin{eqnarray}\label{initialstate}
\ket{\Psi}_{k,k+1}&=&\frac{1}{\sqrt{3}}(e^{-i\pi/4}\ket{gg}-ie^{-i\pi/4}\ket{ee}-i\ket{ff})_{k,k+1}, \qquad   k=1,3,
  \end{eqnarray}
  (states (\ref{initialstate}) are maximally entangled states for two three-level atoms \cite{Zheng2003,Acin2002}, where cavity-assisted collisions of two Rydberg  atoms can be used to produce these Bell states \cite{Zou2003}). The atoms (2,3) with spontaneous emission rates $(\gamma_j,\gamma'_j , j=e,f,g)$ from the $j$-th state interact in a two-mode dissipative cavity field with the photon leakage rates $(\kappa,\kappa')$ which is governed by the Hamiltonian $\hat{H}=\hat{H}_0+\hat{H}_1$ $(\hbar=1)$, where
  \begin{eqnarray}\label{hamiltonian}
  \hat{H}_0&=&\omega\hat{a}^{\dagger}\hat{a}+\omega'\hat{b}^{\dagger}\hat{b}+\sum_{j=e,g,f}(\omega_j \ket{j}_2\bra{j}+\omega'_j \ket{j}_3\bra{j})-i\frac{\kappa}{2}\hat{a}^{\dagger }\hat{a}-i\frac{\kappa'}{2}\hat{b}^{\dagger }\hat{b}\\ \nonumber
  &-&\frac{i}{2}\sum_{j=e,g,f}(\gamma_j \ket{j}_2\bra{j}+\gamma'_j \ket{j}_3\bra{j}),  \\ \nonumber
    \end{eqnarray}
    and
 \begin{eqnarray}
    \hat{H}_1&=&g_1  \left( \hat{a}\ket{e}_2\bra{g}+\hat{a}^{\dagger}\ket{g}_2\bra{e}\right)+g_2 \left( \hat{b}\ket{e}_2\bra{f}+\hat{b}^{\dagger}\ket{f}_2\bra{e}\right)\\ \nonumber
      &+&g'_1  \left( \hat{a}\ket{e}_3\bra{g}+\hat{a}^{\dagger}\ket{g}_3\bra{e}\right)+g'_2 \left( \hat{b}\ket{e}_3\bra{f}+\hat{b}^{\dagger}\ket{f}_3\bra{e}\right),
 \end{eqnarray}
  are the free and interaction parts of Hamiltonian, respectively and $g_l, g'_l$, $l=1,2$ are the atom-field coupling coefficients. Also, $\hat{a}^{\dagger}$ and $\hat{b}^{\dagger}$ $(\hat{a}$ and $\hat{b})$ are creation (annihilation) operators of the two modes of the quantized field and $\omega_j$, $\omega'_j$ ($\omega$, $\omega'$) are the atoms' transitions (fields) frequencies. We assume that $g_1=g'_1$, $g_2=g'_2$ and suppose that the cavities are in vacuum state, so by using the path of Ref. \cite{Zheng2000}, the effective Hamiltonian can be achieved as below:
\begin{eqnarray}\label{effectivehamiltonian}
\hat{H}_{eff}&=&(\lambda_1+\lambda_2)(\ket{e}_2\bra{e}+\ket{e}_3\bra{e})+\lambda_1(\ket{e}_2\bra{g}\ket{g}_3\bra{e}+H.C.)\\ \nonumber
& +&\lambda_2(\ket{e}_2\bra{f}\ket{f}_3\bra{e}+H.C.),\\
\lambda_1&=&\frac{g^2_1}{\Delta-\frac{i}{2}\Gamma},   \qquad \lambda_2=\frac{g^2_2}{\delta-\frac{i}{2}\gamma},\nonumber
\end{eqnarray}
 where we have defined the detuning parameters
 \begin{eqnarray}
     \Delta&=&\omega_e-\omega_g-\omega=\omega'_e-\omega'_g-\omega, \qquad \delta=\omega_e-\omega_f-\omega'=\omega'_e-\omega'_f-\omega',
       \end{eqnarray}
   and the effective atom-field dissipation parameters as
    \begin{eqnarray}\label{effdiss}
     \Gamma&=&\gamma_e-\gamma_g-\kappa=\gamma'_e-\gamma'_g-\kappa,\qquad \gamma=\gamma_e-\gamma_f-\kappa'=\gamma'_e-\gamma'_f-\kappa'.
      \end{eqnarray}
      It is easy to obtain the unnormalized entangled state for atoms $(1-4)$ using the effective Hamiltonian (\ref{effectivehamiltonian}) and initial state (\ref{initialstate}) results in:
\begin{eqnarray}\label{state1-4}
 \ket{\Psi(t)}_{1,2,3,4}&=&(A_1(t)\ket{gggg}+A_2(t)\ket{gege}+A_3(t)\ket{ggee}+A_4(t)\ket{ggff}\\ \nonumber
 &+&A_5(t)\ket{eegg}+A_6(t)\ket{egeg}+A_7(t)\ket{eeee}+A_8(t)\ket{eeff}\\ \nonumber
 &+&A_9(t)\ket{efef}+A_{10}(t)\ket{ffgg}+A_{11}(t)\ket{fefe}+A_{12}(t)\ket{ffee}\\ \nonumber
 &+&A_{13}(t)\ket{ffff})_{1,2,3,4},
\end{eqnarray}
 where
\begin{eqnarray}\label{coefficient}
  A_1(t)&=&\frac{e^{-i\pi/2}}{3}, \\ \nonumber
  A_2(t)&=&A_6(t)=-\frac{e^{-i\pi/2}}{3}e^{-i(\lambda_1+\lambda_2) t}\sin{\lambda_1 t},\\ \nonumber
  A_3(t)&=&A_5(t)=-\frac{ie^{-i\pi/2}}{3}e^{-i(\lambda_1+\lambda_2) t}\cos{\lambda_1 t},\\ \nonumber
  A_4(t)&=&A_{10}=-\frac{ie^{-i\pi/4}}{3}, \qquad A_7(t)=-\frac{e^{-i\pi/2}}{3}e^{-2i(\lambda_1+\lambda_2)t}, \\ \nonumber
  A_8(t)&=&A_{12}(t)=-\frac{e^{-i\pi/4}}{3}e^{-i(\lambda_1+\lambda_2) t}\cos{\lambda_2 t},\\ \nonumber
  A_9(t)&=&A_{11}(t)=\frac{ie^{-i\pi/4}}{3}e^{-i(\lambda_1+\lambda_2) t}\sin{\lambda_2 t}, \qquad  A_{13}(t)=-\frac{1}{3}.
  \end{eqnarray}
   Now, by measuring the atomic states $\ket{eg}_{2,3}$, $\ket{ge}_{2,3}$, $\ket{ef}_{2,3}$ and $\ket{fe}_{2,3}$ on the introduced state in (\ref{state1-4}), the following entangled states for atoms $(1, 4)$ are obtained respectively as
 \begin{eqnarray}\label{state14}
    \ket{\Psi(t)}_{1,4}&=&\frac{1}{\sqrt{\left| A_2(t)\right|^2+\left| A_5(t)\right|^2} }(A_2(t)\ket{ge}+A_5(t)\ket{eg})_{1,4},\\ \nonumber
    \ket{\Psi'(t)}_{1,4}&=&\frac{1}{\sqrt{\left| A_3(t)\right|^2+\left| A_6(t)\right|^2} }(A_3(t)\ket{ge}+A_6(t)\ket{eg})_{1,4},\\ \nonumber
    \ket{\Psi^{\prime\prime}(t)}_{1,4}&=&\frac{1}{\sqrt{\left| A_8(t)\right|^2+\left| A_{11}(t)\right|^2} }(A_8(t)\ket{ef}+A_{11}(t)\ket{fe})_{1,4},\\ \nonumber
     \ket{\Psi^{\prime\prime\prime}(t)}_{1,4}&=&\frac{1}{\sqrt{\left| A_9(t)\right|^2+\left| A_{12}(t)\right|^2} }(A_9(t)\ket{ef}+A_{12}(t)\ket{fe})_{1,4}.
 \end{eqnarray}
 Similar results can be achieved by repeating the above procedure for atoms (5-8) results in:
\begin{eqnarray}\label{state58}
    \ket{\Psi(t)}_{5,8}&=&\frac{1}{\sqrt{\left| A_2(t)\right|^2+\left| A_5(t)\right|^2} }(A_2(t)\ket{ge}+A_5(t)\ket{eg})_{5,8},\\ \nonumber
    \ket{\Psi'(t)}_{5,8}&=&\frac{1}{\sqrt{\left| A_3(t)\right|^2+\left| A_6(t)\right|^2} }(A_3(t)\ket{ge}+A_6(t)\ket{eg})_{5,8},\\ \nonumber
    \ket{\Psi^{\prime\prime}(t)}_{5,8}&=&\frac{1}{\sqrt{\left| A_8(t)\right|^2+\left| A_{11}(t)\right|^2} }(A_8(t)\ket{ef}+A_{11}(t)\ket{fe})_{5,8},\\ \nonumber
    \ket{\Psi^{\prime\prime\prime}(t)}_{5,8}&=&\frac{1}{\sqrt{\left| A_9(t)\right|^2+\left| A_{12}(t)\right|^2} }(A_9(t)\ket{ef}+A_{12}(t)\ket{fe})_{5,8}.
\end{eqnarray}
Now, there exist sixteen possible states for still separable pairs (1,4) and (5,8), where we only consider eight of them with the initial states $\ket{\Psi(t)}_{1,4}\otimes \ket{\Psi(t)}_{5,8}$, $ \ket{\Psi(t)}_{1,4}\otimes \ket{\Psi'(t)}_{5,8}$, $ \ket{\Psi'(t)}_{1,4}\otimes \ket{\Psi(t)}_{5,8}$, $ \ket{\Psi'(t)}_{1,4}\otimes \ket{\Psi'(t)}_{5,8}$, $ \ket{\Psi^{\prime\prime}(t)}_{1,4}\otimes \ket{\Psi^{\prime\prime}(t)}_{5,8}$, $\ket{\Psi^{\prime\prime}(t)}_{1,4}\otimes \ket{\Psi^{\prime\prime\prime}(t)}_{5,8}$, $\ket{\Psi^{\prime\prime\prime}(t)}_{1,4}\otimes \ket{\Psi^{\prime\prime}(t)}_{5,8}$ and $\ket{\Psi^{\prime\prime\prime}(t)}_{1,4}\otimes \ket{\Psi^{\prime\prime\prime}(t)}_{5,8}$. In all cases, by performing the interaction using the dynamical Hamiltonian (\ref{effectivehamiltonian}) between atoms (4,5), the entangled states for atoms (1,8) will be acquired.
\subsection[The first state]{The state $ \ket{\Psi(t)}_{1,4}\otimes \ket{\Psi(t)}_{5,8}$}
The unnormalized entangled state for atoms (1,4,5,8) can be achieved via performing the dissipative interaction using the dynamical Hamiltonian (\ref{effectivehamiltonian}) between atoms (4,5) with initial state $ \ket{\Psi(t)}_{1,4}\otimes \ket{\Psi(t)}_{5,8}$ which is obtained from (\ref{state14}), (\ref{state58}) results in:
\begin{eqnarray}\label{intstate1458}
\small
   \ket{\gamma}^1_{1,4,5,8} & = B^1_1(\tau)\ket{gege}+ B^1_2(\tau)\ket{ggee}+ B^1_3(\tau)\ket{eegg}\\ \nonumber
   &+ B^1_4(\tau)\ket{egeg}+ B^1_5(\tau)\ket{egge}+ B^1_6(\tau)\ket{geeg},
\end{eqnarray}
where
\begin{eqnarray}\label{coefficient11}
  B^1_1(\tau)&=&\frac{A^2_2(t)/2}{\left|A_2(t) \right| ^2+\left|A_5(t) \right| ^2}e^{-i(\lambda_1+\lambda_2) \tau}(e^{i\lambda_1\tau}e^{i\lambda_2 t}+e^{-i\lambda_1\tau} e^{2i\lambda_1 t}e^{i\lambda_2 t}),\\ \nonumber
    B^1_2(\tau)&=&-\frac{A^2_2(t)/2}{\left|A_2(t) \right| ^2+\left|A_5(t) \right| ^2}e^{-i(\lambda_1+\lambda_2) \tau}(e^{i\lambda_1\tau}e^{i\lambda_2 t}-e^{-i\lambda_1\tau} e^{2i\lambda_1 t}e^{i\lambda_2 t}),\\ \nonumber
        B^1_3(\tau)&=&-\frac{A^2_5(t)/2}{\left|A_2(t) \right| ^2+\left|A_5(t) \right| ^2}e^{-i(\lambda_1+\lambda_2) \tau}(e^{i\lambda_1\tau}e^{i\lambda_2 t}-e^{-i\lambda_1\tau} e^{2i\lambda_1 t}e^{i\lambda_2 t}),\\ \nonumber
                B^1_4(\tau)&=&\frac{A^2_5(t)/2}{\left|A_2(t) \right| ^2+\left|A_5(t) \right| ^2}e^{-i(\lambda_1+\lambda_2) \tau}(e^{i\lambda_1\tau}e^{i\lambda_2 t}+e^{-i\lambda_1\tau} e^{2i\lambda_1 t}e^{i\lambda_2 t}),\\ \nonumber
                                B^1_5(\tau)&=&\frac{A_2(t) A_5(t)}{\left|A_2(t) \right| ^2+\left|A_5(t) \right| ^2},\qquad B^1_6(\tau)=\frac{A_2(t) A_5(t)}{\left|A_2(t) \right| ^2+\left|A_5(t) \right| ^2} e^{-2i(\lambda_1+\lambda_2) (\tau-t)}.
 \end{eqnarray}
By measuring the states $\ket{eg}_{4, 5}$ and $\ket{ge}_{4, 5}$ on the state (\ref{intstate1458}), the normalized entangled states of atoms (1,8) respectively collapse to
 \begin{eqnarray}\label{intstate18}
    \ket{\gamma}^1_{1,8}&= &\frac{1}{\sqrt{\left| B^1_1(\tau)\right|^2+\left| B^1_3(\tau)\right|^2 }}\left(B^1_1(\tau)\ket{ge}+B^1_3(\tau)\ket{eg} \right) _{1,8},
  \end{eqnarray}
whose negativity and success probability can be calculated as
\begin{equation}
N_1(\rho)= \frac{\left| B^1_1(\tau) \right| \left| B^1_3(\tau) \right| }{\left| B^1_1(\tau)\right|^2+\left| B^1_3(\tau)\right|^2 }, \qquad S_1(\tau)= \frac{\left| B^1_1(\tau) \right|^2+ \left| B^1_3(\tau) \right|^2 }{\sum^6_{i=1}\left| B^1_i(\tau)\right|^2 },
\end{equation}
and to the state
     \begin{eqnarray}\label{intstate18}
        \ket{\gamma'}^1_{1,8}&= &\frac{1}{\sqrt{\left| B^1_2(\tau)\right|^2+\left| B^1_4(\tau)\right|^2 }}\left(B^1_2(\tau)\ket{ge}+B^1_4(\tau)\ket{eg} \right) _{1,8},
      \end{eqnarray}
    whose negativity and success probability read as
    \begin{equation}
    N'_1(\rho)= \frac{\left| B^1_2(\tau) \right| \left| B^1_4(\tau) \right| }{\left| B^1_2(\tau)\right|^2+\left| B^1_4(\tau)\right|^2 }, \qquad S'_1(\tau)= \frac{\left| B^1_2(\tau) \right|^2+ \left| B^1_4(\tau) \right|^2 }{\sum^6_{i=1}\left| B^1_i(\tau)\right|^2 }.
    \end{equation}
 \subsection[The second state]{The state $ \ket{\Psi(t)}_{1,4}\otimes \ket{\Psi'(t)}_{5,8}$}
By performing the dissipative interaction using the Hamiltonian (\ref{effectivehamiltonian}) between atoms $(4,5)$ with initial state $ \ket{\Psi(t)}_{1,4}\otimes \ket{\Psi'(t)}_{5,8}$ the unnormalized entangled state for atoms $(1,4,5,8)$ may be achieved as
\begin{eqnarray}\label{2intstate}
\small
   \ket{\gamma}^2_{1,4,5,8} & = B^2_1(\tau)\ket{gege}+ B^2_2(\tau)\ket{ggee}+ B^2_3(\tau)\ket{geeg}\\ \nonumber
   &+ B^2_4(\tau)\ket{egge}+ B^2_5(\tau)\ket{eegg}+ B^2_6(\tau)\ket{egeg},
\end{eqnarray}
where
\begin{eqnarray}\label{2coefficient}
  B^2_1(\tau)&=&\frac{A_2(t)A_3(t)e^{-i(\lambda_1+\lambda_2) \tau}}{2\sqrt{\left|A_2(t) \right| ^2+\left|A_5(t) \right| ^2} \sqrt{\left|A_3(t) \right| ^2+\left|A_6(t) \right| ^2}}(e^{i\lambda_1\tau}e^{i\lambda_2 t}+e^{-i\lambda_1\tau} e^{2i\lambda_1 t}e^{i\lambda_2 t}),\\ \nonumber
    B^2_2(\tau)&=&-\frac{A_2(t)A_3(t)e^{-i(\lambda_1+\lambda_2) \tau}}{2\sqrt{\left|A_2(t) \right| ^2+\left|A_5(t) \right| ^2} \sqrt{\left|A_3(t) \right| ^2+\left|A_6(t) \right| ^2}}(e^{i\lambda_1\tau}e^{i\lambda_2 t}-e^{-i\lambda_1\tau} e^{2i\lambda_1 t}e^{i\lambda_2 t}),\\ \nonumber
        B^2_3(\tau)&=&\frac{A_2(t)A_6(t)}{\sqrt{\left|A_2(t) \right| ^2+\left|A_5(t) \right| ^2} \sqrt{\left|A_3(t) \right| ^2+\left|A_6(t) \right| ^2}} e^{-2i(\lambda_1+\lambda_2) (\tau-t)},\\ \nonumber
                B^2_4(\tau)&=&\frac{A_3(t)A_5(t)}{\sqrt{\left|A_2(t) \right| ^2+\left|A_5(t) \right| ^2} \sqrt{\left|A_3(t) \right| ^2+\left|A_6(t) \right| ^2}},\\ \nonumber
                                B^2_5(\tau)&=&-\frac{A_5(t)A_6(t)e^{-i(\lambda_1+\lambda_2) \tau}}{2\sqrt{\left|A_2(t) \right| ^2+\left|A_5(t) \right| ^2} \sqrt{\left|A_3(t) \right| ^2+\left|A_6(t) \right| ^2}}(e^{i\lambda_1\tau}e^{i\lambda_2 t}-e^{-i\lambda_1\tau} e^{2i\lambda_1 t}e^{i\lambda_2 t}),\\ \nonumber
                                 B^2_6(\tau)&=&\frac{A_5(t)A_6(t)e^{-i(\lambda_1+\lambda_2) \tau}}{2\sqrt{\left|A_2(t) \right| ^2+\left|A_5(t) \right| ^2} \sqrt{\left|A_3(t) \right| ^2+\left|A_6(t) \right| ^2}}(e^{i\lambda_1\tau}e^{i\lambda_2 t}+e^{-i\lambda_1\tau} e^{2i\lambda_1 t}e^{i\lambda_2 t}).
 \end{eqnarray}
 Now, by measuring the states $\ket{eg}_{4,5}$ and $\ket{ge}_{4,5}$ on the state (\ref{2intstate}), the normalized entangled states for atoms (1,8) are obtained respectively as
  \begin{eqnarray}\label{2intstate18}
     \ket{\gamma}^2_{1,8}&= &\frac{1}{\sqrt{\left| B^2_1(\tau)\right|^2+\left| B^2_5(\tau)\right|^2 }}\left(B^2_1(\tau)\ket{ge}+B^2_5(\tau)\ket{eg} \right) _{1,8},
   \end{eqnarray}
   with negativity and success probability as
   \begin{equation}\label{2}
 N_2(\rho)= \frac{\left| B^2_1(\tau) \right| \left| B^2_5(\tau) \right| }{\left| B^2_1(\tau)\right|^2+\left| B^2_5(\tau)\right|^2 }, \qquad S_2(\tau)= \frac{\left| B^2_1(\tau) \right|^2+ \left| B^2_5(\tau) \right|^2 }{\sum^6_{i=1}\left| B^2_i(\tau)\right|^2 },
   \end{equation}
   and
 \begin{eqnarray}\label{intstatee2}
 \ket{\gamma'}^2_{1,8}&= &\frac{1}{\sqrt{\left| B^2_2(\tau)\right|^2+\left| B^2_6(\tau)\right|^2 }}\left(B^2_2(\tau)\ket{ge}+B^2_6(\tau)\ket{eg} \right) _{1,8},
\end{eqnarray}
 with negativity and success probability as
\begin{eqnarray}\label{concuu2}
  N'_2(\rho)= \frac{\left| B^2_2(\tau) \right| \left| B^2_6(\tau) \right| }{\left| B^2_2(\tau)\right|^2+\left| B^2_6(\tau)\right|^2 }, \qquad S'_2(\tau)= \frac{\left| B^2_2(\tau) \right|^2+ \left| B^2_6(\tau) \right|^2 }{\sum^6_{i=1}\left| B^2_i(\tau)\right|^2 }.
\end{eqnarray}
Note that $N_2(\rho)=N'_2(\rho)$ and $S_2(\tau)=S'_2(\tau)$.
    \subsection[The third state]{The state $\ket{\Psi'(t)}_{1,4}\otimes \ket{\Psi(t)}_{5,8}$}
 The unnormalized entangled state of atoms $(1,4,5,8)$ is obtained by applying the interaction (\ref{effectivehamiltonian}) on initial state $ \ket{\Psi'(t)}_{1,4}\otimes \ket{\Psi(t)}_{5,8}$ as below:
  \begin{eqnarray}\label{3intstate}
  \small
     \ket{\gamma}^3_{1,4,5,8} & = B^3_1(\tau)\ket{gege}+ B^3_2(\tau)\ket{ggee}+ B^3_3(\tau)\ket{geeg}\\ \nonumber
     &+ B^3_4(\tau)\ket{egge}+ B^3_5(\tau)\ket{eegg}+ B^3_6(\tau)\ket{egeg},
  \end{eqnarray}
  where
  \begin{eqnarray}\label{3coefficient}
    B^3_1(\tau)&=&\frac{A_2(t)A_3(t)e^{-i(\lambda_1+\lambda_2) \tau}}{2\sqrt{\left|A_2(t) \right| ^2+\left|A_5(t) \right| ^2} \sqrt{\left|A_3(t) \right| ^2+\left|A_6(t) \right| ^2}}(e^{i\lambda_1\tau}e^{i\lambda_2 t}+e^{-i\lambda_1\tau} e^{2i\lambda_1 t}e^{i\lambda_2 t}),\\ \nonumber
      B^3_2(\tau)&=&-\frac{A_2(t)A_3(t)e^{-i(\lambda_1+\lambda_2) \tau}}{2\sqrt{\left|A_2(t) \right| ^2+\left|A_5(t) \right| ^2} \sqrt{\left|A_3(t) \right| ^2+\left|A_6(t) \right| ^2}}(e^{i\lambda_1\tau}e^{i\lambda_2 t}-e^{-i\lambda_1\tau} e^{2i\lambda_1 t}e^{i\lambda_2 t}),\\ \nonumber
          B^3_3(\tau)&=&\frac{A_3(t)A_5(t)}{\sqrt{\left|A_2(t) \right| ^2+\left|A_5(t) \right| ^2} \sqrt{\left|A_3(t) \right| ^2+\left|A_6(t) \right| ^2}} e^{-2i(\lambda_1+\lambda_2) (\tau-t)},\\ \nonumber
                  B^3_4(\tau)&=&\frac{A_2(t)A_6(t)}{\sqrt{\left|A_2(t) \right| ^2+\left|A_5(t) \right| ^2} \sqrt{\left|A_3(t) \right| ^2+\left|A_6(t) \right| ^2}},\\ \nonumber
                                  B^3_5(\tau)&=&-\frac{A_5(t)A_6(t)e^{-i(\lambda_1+\lambda_2) \tau}}{2\sqrt{\left|A_2(t) \right| ^2+\left|A_5(t) \right| ^2} \sqrt{\left|A_3(t) \right| ^2+\left|A_6(t) \right| ^2}}(e^{i\lambda_1\tau}e^{i\lambda_2 t}-e^{-i\lambda_1\tau} e^{2i\lambda_1 t}e^{i\lambda_2 t}),\\ \nonumber
                                   B^3_6(\tau)&=&\frac{A_5(t)A_6(t)e^{-i(\lambda_1+\lambda_2) \tau}}{2\sqrt{\left|A_2(t) \right| ^2+\left|A_5(t) \right| ^2} \sqrt{\left|A_3(t) \right| ^2+\left|A_6(t) \right| ^2}}(e^{i\lambda_1\tau}e^{i\lambda_2 t}+e^{-i\lambda_1\tau} e^{2i\lambda_1 t}e^{i\lambda_2 t}).
   \end{eqnarray}
       By measuring the states $\ket{eg}_{4,5}$ and $\ket{ge}_{4,5}$ on the state (\ref{3intstate}), the entangled states of atoms $(1,8)$ are respectively converted to $\ket{\gamma}^3_{1,8}$ and $\ket{\gamma'}^3_{1,8}$, where $\ket{\gamma}^3_{1,8}= \ket{\gamma}^2_{1,8}$ and $\ket{\gamma'}^3_{1,8}= \ket{\gamma'}^2_{1,8}$ respectively introduced in (\ref{2intstate18}) and (\ref{intstatee2}), with calculated negativities $N_3(\rho)=N_2(\rho)$, $N'_3(\rho)=N'_2(\rho)$ and success probabilities\\
  \begin{eqnarray}\label{success3}
    S_3(\rho)= \frac{\left| B^3_1(\tau) \right|^2+ \left| B^3_5(\tau) \right|^2 }{\sum^6_{i=1}\left| B^3_i(\tau)\right|^2 }, \qquad S'_3(\tau)= \frac{\left| B^3_2(\tau) \right|^2+ \left| B^3_6(\tau) \right|^2 }{\sum^6_{i=1}\left| B^3_i(\tau)\right|^2 }.
  \end{eqnarray}
        Note that $N_2(\rho)=N'_2(\rho)=N_3(\rho)=N'_3(\rho)$ and $S_3(\tau)=S'_3(\tau)$ respectively introduced in (\ref{2}), (\ref{concuu2}) and (\ref{success3}).
   \subsection[The forth state]{The state $ \ket{\Psi'(t)}_{1,4}\otimes \ket{\Psi'(t)}_{5,8}$}
      The above procedure is now repeated for initial state $ \ket{\Psi'(t)}_{1,4}\otimes \ket{\Psi'(t)}_{5,8}$, so the unnormalized entangled state for atoms (1,4,5,8) is achieved as below:
  \begin{eqnarray}\label{4intstate}
   \small
      \ket{\gamma}^4_{1,4,5,8} & = B^4_1(\tau)\ket{gege}+ B^4_2(\tau)\ket{ggee}+ B^4_3(\tau)\ket{eegg}\\ \nonumber
      &+ B^4_4(\tau)\ket{egeg}+ B^4_5(\tau)\ket{geeg}+ B^4_6(\tau)\ket{egge},
   \end{eqnarray}
   where
   \begin{eqnarray}\label{4coefficient}
    B^4_1(\tau)&=&\frac{A^2_3(t)/2}{\left|A_3(t) \right| ^2+\left|A_6(t) \right| ^2}e^{-i(\lambda_1+\lambda_2) \tau}(e^{i\lambda_1\tau}e^{i\lambda_2 t}+e^{-i\lambda_1\tau} e^{2i\lambda_1 t}e^{i\lambda_2 t}),\\ \nonumber
       B^4_2(\tau)&=&-\frac{A^2_3(t)/2}{\left|A_3(t) \right| ^2+\left|A_6(t) \right| ^2}e^{-i(\lambda_1+\lambda_2) \tau}(e^{i\lambda_1\tau}e^{i\lambda_2 t}-e^{-i\lambda_1\tau} e^{2i\lambda_1 t}e^{i\lambda_2 t}),\\ \nonumber
           B^4_3(\tau)&=&-\frac{A^2_6(t)/2}{\left|A_3(t) \right| ^2+\left|A_6(t) \right| ^2}e^{-i(\lambda_1+\lambda_2) \tau}(e^{i\lambda_1\tau}e^{i\lambda_2 t}-e^{-i\lambda_1\tau} e^{2i\lambda_1 t}e^{i\lambda_2 t}),\\ \nonumber
                   B^4_4(\tau)&=&\frac{A^2_6(t)/2}{\left|A_3(t) \right| ^2+\left|A_6(t) \right| ^2}e^{-i(\lambda_1+\lambda_2) \tau}(e^{i\lambda_1\tau}e^{i\lambda_2 t}+e^{-i\lambda_1\tau} e^{2i\lambda_1 t}e^{i\lambda_2 t}),\\ \nonumber
                                   B^4_5(\tau)&=&\frac{A_3(t) A_6(t)}{\left|A_3(t) \right| ^2+\left|A_6(t) \right| ^2}e^{-2i(\lambda_1+\lambda_2) (\tau-t)},\qquad B^4_6(\tau)=\frac{A_3(t) A_6(t)}{\left|A_3(t) \right| ^2+\left|A_6(t) \right| ^2} .
    \end{eqnarray}
   By measuring the states $\ket{eg}_{4,5}$ and $\ket{ge}_{4,5}$ on the state (\ref{4intstate}), the normalized entangled states of atoms (1,8) can be achieved respectively as,
   \begin{eqnarray}\label{4intstatee}
      \ket{\gamma}^4_{1,8}&= &\frac{1}{\sqrt{\left| B^4_1(\tau)\right|^2+\left| B^4_3(\tau)\right|^2 }}\left(B^4_1(\tau)\ket{ge}+B^4_3(\tau)\ket{eg} \right) _{1,8},
    \end{eqnarray}
  whose negativity and success probability read as
  \begin{equation}
N_4(\rho)= \frac{\left| B^4_1(\tau) \right| \left| B^4_3(\tau) \right| }{\left| B^4_1(\tau)\right|^2+\left| B^4_3(\tau)\right|^2 }, \qquad S_4(\tau)= \frac{\left| B^4_1(\tau) \right|^2+ \left| B^4_3(\tau) \right|^2 }{\sum^6_{i=1}\left| B^4_i(\tau)\right|^2 },
  \end{equation}
and
   \begin{eqnarray}\label{intstatee4}
       \ket{\gamma'}^4_{1,8}&= &\frac{1}{\sqrt{\left| B^4_2(\tau)\right|^2+\left| B^4_4(\tau)\right|^2 }}\left(B^4_2(\tau)\ket{ge}+B^4_4(\tau)\ket{eg} \right) _{1,8},
    \end{eqnarray}
  whose negativity and success probability read as
  \begin{equation}
N'_4(\rho)= \frac{\left| B^4_2(\tau) \right| \left| B^4_4(\tau) \right| }{\left| B^4_2(\tau)\right|^2+\left| B^4_4(\tau)\right|^2 }, \qquad S'_4(\tau)= \frac{\left| B^4_2(\tau) \right|^2+ \left| B^4_4(\tau) \right|^2 }{\sum^6_{i=1}\left| B^4_i(\tau)\right|^2 }.
  \end{equation}
  Note that $N_4(\rho)=N'_1(\rho)$, $N'_4(\rho)=N_1(\rho)$, $S_4(\tau)=S'_1(\tau)$ and $S'_4(\tau)=S_1(\tau)$.
 \subsection[The fifth state]{The state $ \ket{\Psi^{\prime\prime}(t)}_{1,4}\otimes \ket{\Psi^{\prime\prime}(t)}_{5,8}$}
      Using the interaction Hamiltonian (\ref{effectivehamiltonian}) between atoms (4,5) with initial state $ \ket{\Psi^{\prime\prime}(t)}_{1,4}\otimes \ket{\Psi^{\prime\prime}(t)}_{5,8}$, the following unnormalized entangled state for atoms (1,4,5,8) is achieved,
  \begin{eqnarray}\label{5intstate}
   \small
      \ket{\gamma}^5_{1,4,5,8} & = B^5_1(\tau)\ket{eeff}+ B^5_2(\tau)\ket{efef}+ B^5_3(\tau)\ket{ffee}\\ \nonumber
      &+ B^5_4(\tau)\ket{fefe}+ B^5_5(\tau)\ket{effe}+ B^5_6(\tau)\ket{feef},
   \end{eqnarray}
   where
   \begin{eqnarray}\label{5coefficient}
    B^5_1(\tau)&=&-\frac{A^2_8(t)/2}{\left|A_8(t) \right| ^2+\left|A_{11}(t) \right| ^2}e^{-i(\lambda_1+\lambda_2) \tau}(e^{i\lambda_2\tau}e^{i\lambda_1 t}-e^{-i\lambda_2\tau} e^{i\lambda_1 t}e^{2i\lambda_2 t}),\\ \nonumber
       B^5_2(\tau)&=&\frac{A^2_8(t)/2}{\left|A_8(t) \right| ^2+\left|A_{11}(t) \right| ^2}e^{-i(\lambda_1+\lambda_2) \tau}(e^{i\lambda_2\tau}e^{i\lambda_1 t}+e^{-i\lambda_2\tau} e^{i\lambda_1 t}e^{2i\lambda_2 t}),\\ \nonumber
           B^5_3(\tau)&=&-\frac{A^2_{11}(t)/2}{\left|A_8(t) \right| ^2+\left|A_{11}(t) \right| ^2}e^{-i(\lambda_1+\lambda_2) \tau}(e^{i\lambda_2\tau}e^{i\lambda_1 t}-e^{-i\lambda_2\tau} e^{i\lambda_1 t}e^{2i\lambda_2 t}),\\ \nonumber
                   B^5_4(\tau)&=&\frac{A^2_{11}(t)/2}{\left|A_8(t) \right| ^2+\left|A_{11}(t) \right| ^2}e^{-i(\lambda_1+\lambda_2) \tau}(e^{i\lambda_2\tau}e^{i\lambda_1 t}+e^{-i\lambda_2\tau} e^{i\lambda_1 t}e^{2i\lambda_2 t}),\\ \nonumber
                                   B^5_5(\tau)&=&\frac{A_8(t) A_{11}(t)}{\left|A_8(t) \right| ^2+\left|A_{11}(t) \right| ^2},\qquad B^5_6(\tau)=\frac{A_8(t) A_{11}(t)}{\left|A_8(t) \right| ^2+\left|A_{11}(t) \right| ^2}e^{-2i(\lambda_1+\lambda_2) (\tau-t)} .
    \end{eqnarray}
   By measuring the states $\ket{ef}_{4,5}$ and $\ket{fe}_{4,5}$ on the state (\ref{5intstate}), the normalized entangled states of atoms (1,8), respectively collapse to,
   \begin{eqnarray}\label{5intstate18}
      \ket{\gamma}^5_{1,8}&= &\frac{1}{\sqrt{\left| B^5_1(\tau)\right|^2+\left| B^5_4(\tau)\right|^2 }}\left(B^5_1(\tau)\ket{ef}+B^5_4(\tau)\ket{fe} \right) _{1,8},
    \end{eqnarray}
  with negativity and success probability as
  \begin{equation}
N_5(\rho)= \frac{\left| B^5_1(\tau) \right| \left| B^5_4(\tau) \right| }{\left| B^5_1(\tau)\right|^2+\left| B^5_4(\tau)\right|^2 }, \qquad S_5(\tau)= \frac{\left| B^5_1(\tau) \right|^2+ \left| B^5_4(\tau) \right|^2 }{\sum^6_{i=1}\left| B^5_i(\tau)\right|^2 },
  \end{equation}
and
   \begin{eqnarray}\label{intstatee184}
       \ket{\gamma'}^5_{1,8}&= &\frac{1}{\sqrt{\left| B^5_2(\tau)\right|^2+\left| B^5_3(\tau)\right|^2 }}\left(B^5_2(\tau)\ket{ef}+B^5_3(\tau)\ket{fe} \right) _{1,8},
    \end{eqnarray}
  with negativity and success probability as,
  \begin{equation}
N'_5(\rho)= \frac{\left| B^5_2(\tau) \right| \left| B^5_3(\tau) \right| }{\left| B^5_2(\tau)\right|^2+\left| B^5_3(\tau)\right|^2 }, \qquad S'_5(\tau)= \frac{\left| B^5_2(\tau) \right|^2+ \left| B^5_3(\tau) \right|^2 }{\sum^6_{i=1}\left| B^5_i(\tau)\right|^2 }.
  \end{equation}
   \subsection[The sixth state]{The state $\ket{\Psi^{\prime\prime}(t)}_{1,4}\otimes \ket{\Psi^{\prime\prime\prime}(t)}_{5,8}$}
  By performing the interaction Hamiltonian (\ref{effectivehamiltonian}) between atoms $(4,5)$ with initial state $\ket{\Psi^{\prime\prime}(t)}_{1,4}\otimes \ket{\Psi^{\prime\prime\prime}(t)}_{5,8}$, the unnormalized entangled state for atoms $(1,4,5,8)$ can be achieved as
  \begin{eqnarray}\label{6intstate}
  \small
     \ket{\gamma}^6_{1,4,5,8} & = B^6_1(\tau)\ket{efef}+ B^6_2(\tau)\ket{eeff}+ B^6_3(\tau)\ket{effe}\\ \nonumber
     &+ B^6_4(\tau)\ket{feef}+ B^6_5(\tau)\ket{fefe}+ B^6_6(\tau)\ket{ffee},
  \end{eqnarray}
  where
  \begin{eqnarray}\label{coefficient12}
    B^6_1(\tau)&=&\frac{A_8(t)A_9(t)e^{-i(\lambda_1+\lambda_2) \tau}}{2\sqrt{\left|A_8(t) \right| ^2+\left|A_{11}(t) \right| ^2} \sqrt{\left|A_9(t) \right| ^2+\left|A_{12}(t) \right| ^2}}(e^{i\lambda_2\tau}e^{i\lambda_1 t}+e^{-i\lambda_2\tau} e^{i\lambda_1 t}e^{2i\lambda_2 t}),\\ \nonumber
      B^6_2(\tau)&=&-\frac{A_8(t)A_9(t)e^{-i(\lambda_1+\lambda_2) \tau}}{2\sqrt{\left|A_8(t) \right| ^2+\left|A_{11}(t) \right| ^2} \sqrt{\left|A_9(t) \right| ^2+\left|A_{12}(t) \right| ^2}}(e^{i\lambda_2\tau}e^{i\lambda_1 t}-e^{-i\lambda_2\tau} e^{i\lambda_1 t}e^{2i\lambda_2 t}),\\ \nonumber
          B^6_3(\tau)&=&\frac{A_8(t)A_{12}(t)}{\sqrt{\left|A_8(t) \right| ^2+\left|A_{11}(t) \right| ^2} \sqrt{\left|A_9(t) \right| ^2+\left|A_{12}(t) \right| ^2}},\\ \nonumber
                            B^6_4(\tau)&=&\frac{A_9(t)A_{11}(t)}{\sqrt{\left|A_8(t) \right| ^2+\left|A_{11}(t) \right| ^2} \sqrt{\left|A_9(t) \right| ^2+\left|A_{12}(t) \right| ^2}}e^{-2i(\lambda_1+\lambda_2)(\tau-t)},\\ \nonumber
                                  B^6_5(\tau)&=&\frac{A_{11}(t)A_{12}(t)e^{-i(\lambda_1+\lambda_2) \tau}}{2\sqrt{\left|A_8(t) \right| ^2+\left|A_{11}(t) \right| ^2} \sqrt{\left|A_9(t) \right| ^2+\left|A_{12}(t) \right| ^2}}(e^{i\lambda_2\tau}e^{i\lambda_1 t}+e^{-i\lambda_2\tau} e^{i\lambda_1 t}e^{2i\lambda_2 t}),\\ \nonumber
                                   B^6_6(\tau)&=&-\frac{A_{11}(t)A_{12}(t)e^{-i(\lambda_1+\lambda_2) \tau}}{2\sqrt{\left|A_8(t) \right| ^2+\left|A_{11}(t) \right| ^2} \sqrt{\left|A_9(t) \right| ^2+\left|A_{12}(t) \right| ^2}}(e^{i\lambda_2\tau}e^{i\lambda_1 t}-e^{-i\lambda_2\tau} e^{i\lambda_1 t}e^{2i\lambda_2 t}).
   \end{eqnarray}
   By measuring the states $\ket{ef}_{4,5}$ and $\ket{fe}_{4,5}$ on state (\ref{6intstate}), the normalized entangled states of atoms (1,8) are respectively achieved as,
    \begin{eqnarray}\label{66intstate182}
       \ket{\gamma}^6_{1,8}&= &\frac{1}{\sqrt{\left| B^6_2(\tau)\right|^2+\left| B^6_5(\tau)\right|^2 }}\left(B^6_2(\tau)\ket{ef}+B^6_5(\tau)\ket{fe} \right) _{1,8},
     \end{eqnarray}
     with negativity and success probability as
     \begin{equation}\label{n6}
   N_6(\rho)= \frac{\left| B^6_2(\tau) \right| \left| B^6_5(\tau) \right| }{\left| B^6_2(\tau)\right|^2+\left| B^6_5(\tau)\right|^2 }, \qquad S_6(\tau)= \frac{\left| B^6_2(\tau) \right|^2+ \left| B^6_5(\tau) \right|^2 }{\sum^6_{i=1}\left| B^6_i(\tau)\right|^2 },
     \end{equation}
     and
   \begin{eqnarray}\label{66intstatee182}
   \ket{\gamma'}^6_{1,8}&= &\frac{1}{\sqrt{\left| B^6_1(\tau)\right|^2+\left| B^6_6(\tau)\right|^2 }}\left(B^6_1(\tau)\ket{ef}+B^6_6(\tau)\ket{fe} \right) _{1,8},
  \end{eqnarray}
   with negativity and success probability as
  \begin{eqnarray}\label{nn6}
    N'_6(\rho)= \frac{\left| B^6_1(\tau) \right| \left| B^6_6(\tau) \right| }{\left| B^6_1(\tau)\right|^2+\left| B^6_6(\tau)\right|^2 }, \qquad S'_6(\tau)= \frac{\left| B^6_1(\tau) \right|^2+ \left| B^6_6(\tau) \right|^2 }{\sum^6_{i=1}\left| B^6_i(\tau)\right|^2 }.
  \end{eqnarray}
  Note that $N_6(\rho)=N'_6(\rho)$ and $S_6(\tau)=S'_6(\tau)$.
    \subsection[The seventh state]{The state $\ket{\Psi^{\prime\prime\prime}(t)}_{1,4}\otimes \ket{\Psi^{\prime\prime}(t)}_{5,8}$}
    By performing the interaction Hamiltonian (\ref{effectivehamiltonian}) between atoms $(4,5)$ with initial state $\ket{\Psi^{\prime\prime\prime}(t)}_{1,4}\otimes \ket{\Psi^{\prime\prime}(t)}_{5,8}$, the unnormalized entangled state for atoms $(1,4,5,8)$ can be deduced as
    \begin{eqnarray}\label{7intstate}
    \small
       \ket{\gamma}^7_{1,4,5,8} & = B^7_1(\tau)\ket{efef}+ B^7_2(\tau)\ket{eeff}+ B^7_3(\tau)\ket{effe}\\ \nonumber
       &+ B^7_4(\tau)\ket{feef}+ B^7_5(\tau)\ket{fefe}+ B^7_6(\tau)\ket{ffee},
    \end{eqnarray}
    where
    \begin{eqnarray}\label{7coefficient}
      B^7_1(\tau)&=&\frac{A_8(t)A_9(t)e^{-i(\lambda_1+\lambda_2) \tau}}{2\sqrt{\left|A_8(t) \right| ^2+\left|A_{11}(t) \right| ^2} \sqrt{\left|A_9(t) \right| ^2+\left|A_{12}(t) \right| ^2}}(e^{i\lambda_2\tau}e^{i\lambda_1 t}+e^{-i\lambda_2\tau} e^{i\lambda_1 t}e^{2i\lambda_2 t}),\\ \nonumber
        B^7_2(\tau)&=&-\frac{A_8(t)A_9(t)e^{-i(\lambda_1+\lambda_2) \tau}}{2\sqrt{\left|A_8(t) \right| ^2+\left|A_{11}(t) \right| ^2} \sqrt{\left|A_9(t) \right| ^2+\left|A_{12}(t) \right| ^2}}(e^{i\lambda_2\tau}e^{i\lambda_1 t}-e^{-i\lambda_2\tau} e^{i\lambda_1 t}e^{2i\lambda_2 t}),\\ \nonumber
            B^7_3(\tau)&=&\frac{A_9(t)A_{11}(t)}{\sqrt{\left|A_8(t) \right| ^2+\left|A_{11}(t) \right| ^2} \sqrt{\left|A_9(t) \right| ^2+\left|A_{12}(t) \right| ^2}},\\ \nonumber
                    B^7_4(\tau)&=&\frac{A_8(t)A_{12}(t)}{\sqrt{\left|A_8(t) \right| ^2+\left|A_{11}(t) \right| ^2} \sqrt{\left|A_9(t) \right| ^2+\left|A_{12}(t) \right| ^2}}e^{-2i(\lambda_1+\lambda_2)(\tau-t)} ,\\ \nonumber
                                    B^7_5(\tau)&=&\frac{A_{11}(t)A_{12}(t)e^{-i(\lambda_1+\lambda_2) \tau}}{2\sqrt{\left|A_8(t) \right| ^2+\left|A_{11}(t) \right| ^2} \sqrt{\left|A_9(t) \right| ^2+\left|A_{12}(t) \right| ^2}}(e^{i\lambda_2\tau}e^{i\lambda_1 t}+e^{-i\lambda_2\tau} e^{i\lambda_1 t}e^{2i\lambda_2 t}),\\ \nonumber
                                     B^7_6(\tau)&=&-\frac{A_{11}(t)A_{12}(t)e^{-i(\lambda_1+\lambda_2) \tau}}{2\sqrt{\left|A_8(t) \right| ^2+\left|A_{11}(t) \right| ^2} \sqrt{\left|A_9(t) \right| ^2+\left|A_{12}(t) \right| ^2}}(e^{i\lambda_2\tau}e^{i\lambda_1 t}-e^{-i\lambda_2\tau} e^{i\lambda_1 t}e^{2i\lambda_2 t}).
     \end{eqnarray}
     Now, by measuring the states $\ket{ef}_{4,5}$ and $\ket{fe}_{4,5}$ on the state (\ref{7intstate}), the entangled states of atoms $(1,8)$ are respectively converted to $\ket{\gamma}^7_{1,8}$ and $\ket{\gamma'}^7_{1,8}$, where $\ket{\gamma}^7_{1,8}= \ket{\gamma}^6_{1,8}$ and $\ket{\gamma'}^7_{1,8}= \ket{\gamma'}^6_{1,8}$ introduced in (\ref{66intstate182}) and (\ref{66intstatee182}), with negativities $N_7(\rho)=N_6(\rho)$, $N'_7(\rho)=N'_6(\rho)$ respectively introduced in (\ref{n6}), (\ref{nn6}), and success probabilities\\
      \begin{eqnarray}
        S_7(\rho)= \frac{\left| B^7_2(\tau) \right|^2+ \left| B^7_5(\tau) \right|^2 }{\sum^6_{i=1}\left| B^7_i(\tau)\right|^2 }, \qquad S'_7(\tau)= \frac{\left| B^7_1(\tau) \right|^2+ \left| B^7_6(\tau) \right|^2 }{\sum^6_{i=1}\left| B^7_i(\tau)\right|^2 }.
      \end{eqnarray}
     Note that $N_7(\rho)=N'_7(\rho)=N_6(\rho)=N'_6(\rho)$ and $S_7(\tau)=S'_7(\tau)$.
     \subsection[The eighth state]{The state $\ket{\Psi^{\prime\prime\prime}(t)}_{1,4}\otimes \ket{\Psi^{\prime\prime\prime}(t)}_{5,8}$}
     The unnormalized entangled state for atoms (1,4,5,8) is achieved by performing the dissipative interaction (\ref{effectivehamiltonian}) between atoms (4,5) with initial state $\ket{\Psi^{\prime\prime\prime}(t)}_{1,4}\otimes \ket{\Psi^{\prime\prime\prime}(t)}_{5,8}$ as,
     \begin{eqnarray}\label{8intstate}
     \small
        \ket{\gamma}^8_{1,4,5,8} & = B^8_1(\tau)\ket{efef}+ B^8_2(\tau)\ket{eeff}+ B^8_3(\tau)\ket{fefe}\\ \nonumber
        &+ B^8_4(\tau)\ket{ffee}+ B^8_5(\tau)\ket{effe}+ B^8_6(\tau)\ket{feef},
     \end{eqnarray}
     where
     \begin{eqnarray}\label{8coefficient}
       B^8_1(\tau)&=&\frac{A^2_9(t)/2}{\left|A_9(t) \right| ^2+\left|A_{12}(t) \right| ^2}e^{-i(\lambda_1+\lambda_2) \tau}(e^{i\lambda_2\tau}e^{i\lambda_1 t}+e^{-i\lambda_2\tau} e^{2i\lambda_2 t}e^{i\lambda_1 t}),\\ \nonumber
         B^8_2(\tau)&=&-\frac{A^2_9(t)/2}{\left|A_9(t) \right| ^2+\left|A_{12}(t) \right| ^2}e^{-i(\lambda_1+\lambda_2) \tau}(e^{i\lambda_2\tau}e^{i\lambda_1 t}-e^{-i\lambda_2\tau} e^{2i\lambda_2 t}e^{i\lambda_1 t}),\\ \nonumber
             B^8_3(\tau)&=&\frac{A^2_{12}(t)/2}{\left|A_9(t) \right| ^2+\left|A_{12}(t) \right| ^2}e^{-i(\lambda_1+\lambda_2) \tau}(e^{i\lambda_2\tau}e^{i\lambda_1 t}+e^{-i\lambda_2\tau} e^{2i\lambda_2 t}e^{i\lambda_1 t}),\\ \nonumber
                     B^8_4(\tau)&=&-\frac{A^2_{12}(t)/2}{\left|A_9(t) \right| ^2+\left|A_{12}(t) \right| ^2}e^{-i(\lambda_1+\lambda_2) \tau}(e^{i\lambda_2\tau}e^{i\lambda_1 t}-e^{-i\lambda_2\tau} e^{2i\lambda_2 t}e^{i\lambda_1 t}),\\ \nonumber
                                     B^8_5(\tau)&=&\frac{A_9(t) A_{12}(t)}{\left|A_9(t) \right| ^2+\left|A_{12}(t) \right| ^2},\qquad B^8_6(\tau)=\frac{A_9(t) A_{12}(t)}{\left|A_9(t) \right| ^2+\left|A_{12}(t) \right| ^2} e^{-2i(\lambda_1+\lambda_2) (\tau-t)}.
      \end{eqnarray}
     By measuring the states $\ket{ef}_{4, 5}$ and $\ket{fe}_{4, 5}$ on the state (\ref{8intstate}), the normalized entangled states of atoms (1,8), respectively collapse to the state,
      \begin{eqnarray}\label{8intstate18}
         \ket{\gamma}^8_{1,8}&= &\frac{1}{\sqrt{\left| B^8_2(\tau)\right|^2+\left| B^8_3(\tau)\right|^2 }}\left(B^8_2(\tau)\ket{ef}+B^8_3(\tau)\ket{fe} \right)_{1,8},
       \end{eqnarray}
     which possesses the negativity and success probability as
     \begin{equation}
     N_8(\rho)= \frac{\left| B^8_2(\tau) \right| \left| B^8_3(\tau) \right| }{\left| B^8_2(\tau)\right|^2+\left| B^8_3(\tau)\right|^2 }, \qquad S_8(\tau)= \frac{\left| B^8_2(\tau) \right|^2+ \left| B^8_3(\tau) \right|^2 }{\sum^6_{i=1}\left| B^8_i(\tau)\right|^2 },
     \end{equation}
and to the state,
          \begin{eqnarray}\label{8intstatee18}
             \ket{\gamma'}^8_{1,8}&= &\frac{1}{\sqrt{\left| B^8_1(\tau)\right|^2+\left| B^8_4(\tau)\right|^2 }}\left(B^8_1(\tau)\ket{ef}+B^8_4(\tau)\ket{fe} \right) _{1,8},
           \end{eqnarray}
         with negativity and success probability as,
         \begin{equation}
         N'_8(\rho)= \frac{\left| B^8_1(\tau) \right| \left| B^8_4(\tau) \right| }{\left| B^8_1(\tau)\right|^2+\left| B^8_4(\tau)\right|^2 }, \qquad S'_8(\tau)= \frac{\left| B^8_1(\tau) \right|^2+ \left| B^8_4(\tau) \right|^2 }{\sum^6_{i=1}\left| B^8_i(\tau)\right|^2 },
         \end{equation}
         where $N_8(\rho)=N'_5(\rho)$, $N'_8(\rho)=N_5(\rho)$, $S_8(\tau)=S'_5(\tau)$ and $S'_8(\tau)=S_5(\tau)$.
 \section{Results and discussion} \label{sec.results}
\begin{figure}
 \centering
 \subfigure[\label{fig.Fig2a} \ $N_1(\rho)=N'_4(\rho)$]{\includegraphics[width=0.35\textwidth]{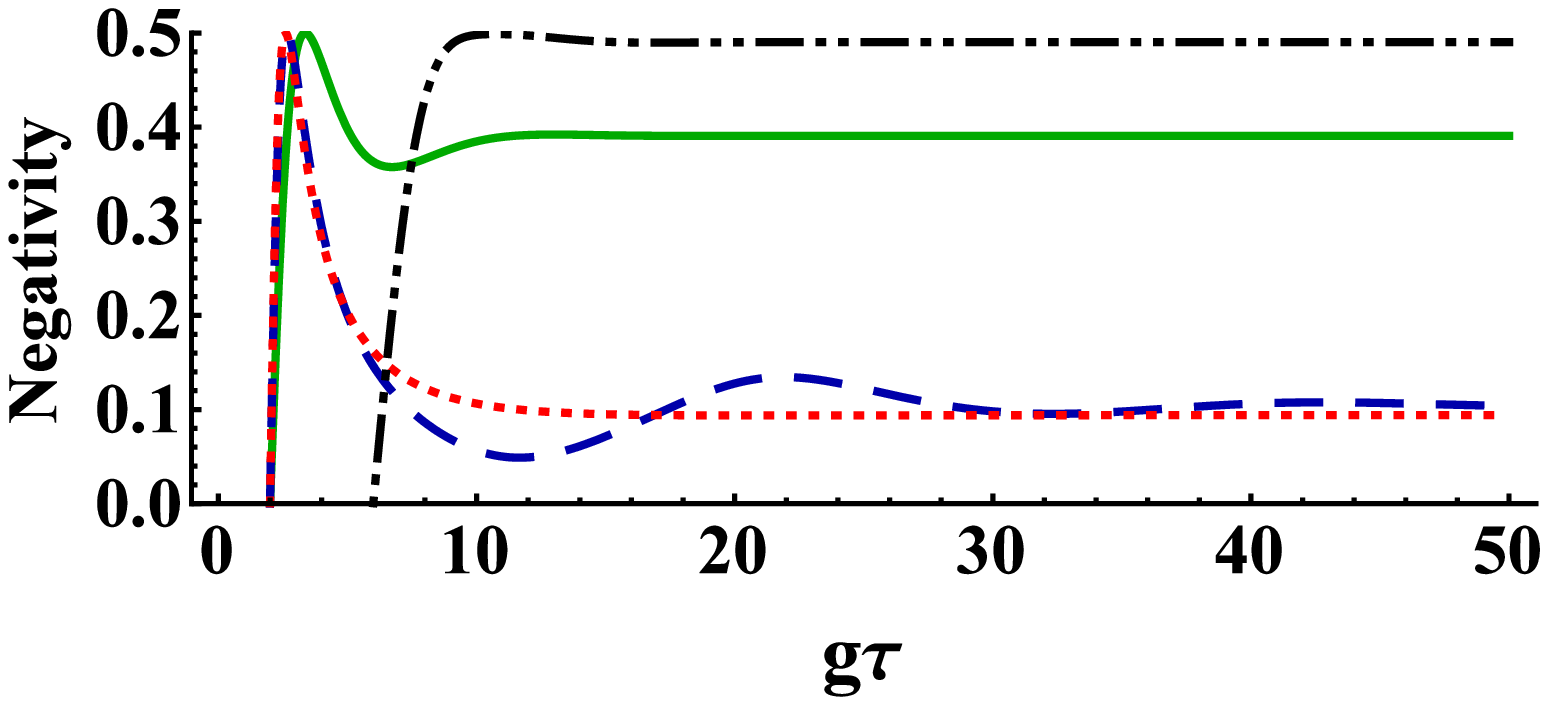}}
 \hspace{0.05\textwidth}
  \subfigure[\label{fig.Fig2b} \ $N'_1(\rho)=N_4(\rho)$]{\includegraphics[width=0.35\textwidth]{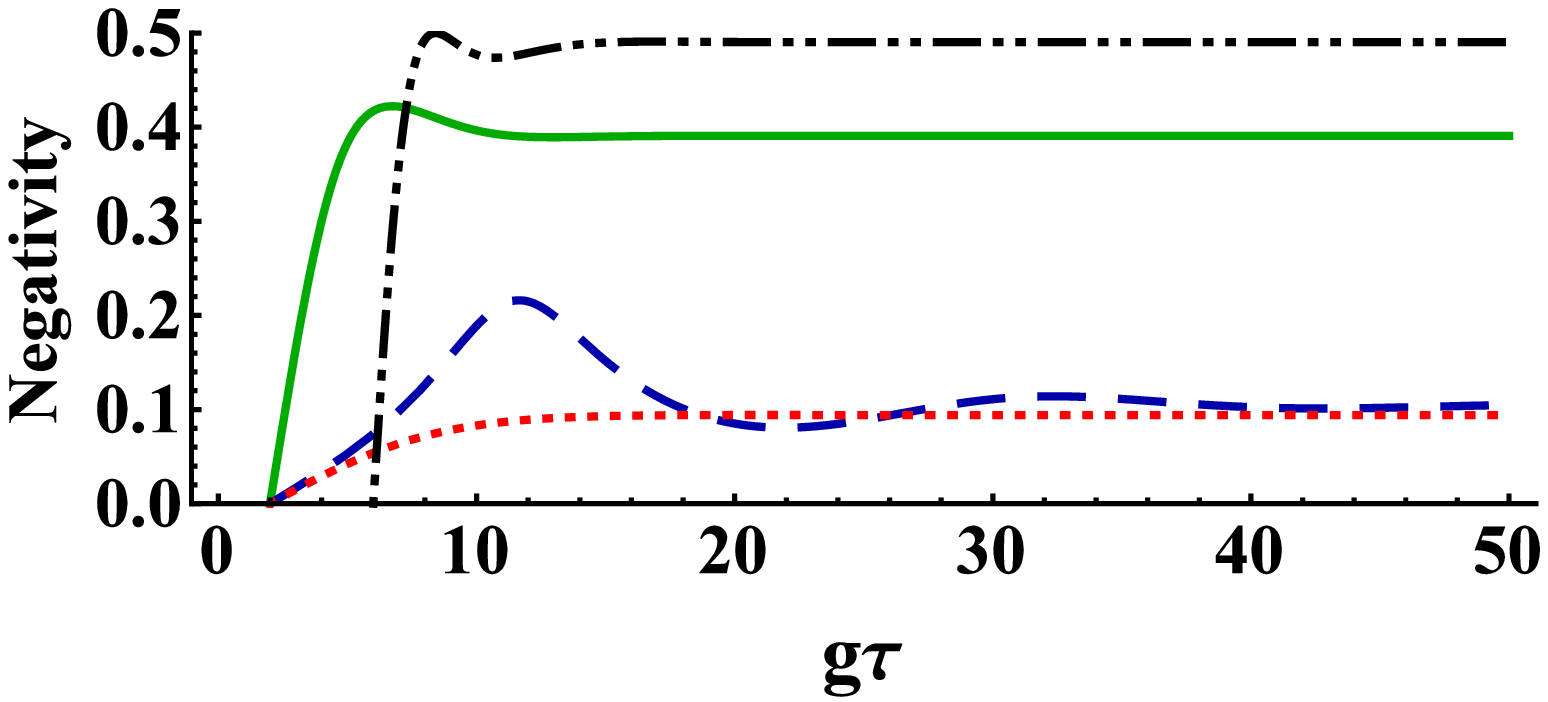}}
  \hspace{0.05\textwidth}
   \subfigure[\label{fig.Fig2c} \ $N_2(\rho)=N'_2(\rho)=N_3(\rho)=N'_3(\rho)$]{\includegraphics[width=0.35\textwidth]{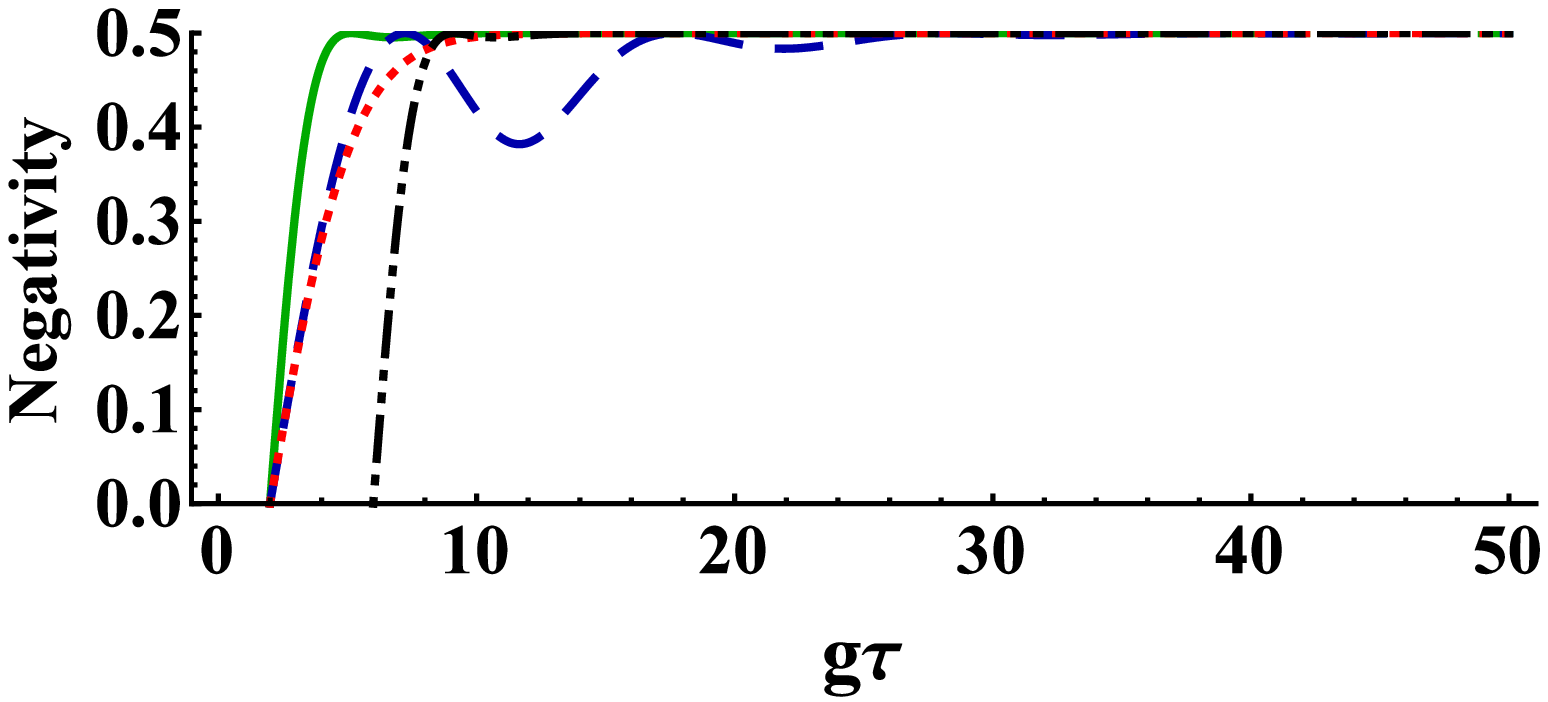}}
   \hspace{0.05\textwidth}
    \subfigure[\label{fig.Fig2d} \ $N_5(\rho)=N'_8(\rho)$]{\includegraphics[width=0.35\textwidth]{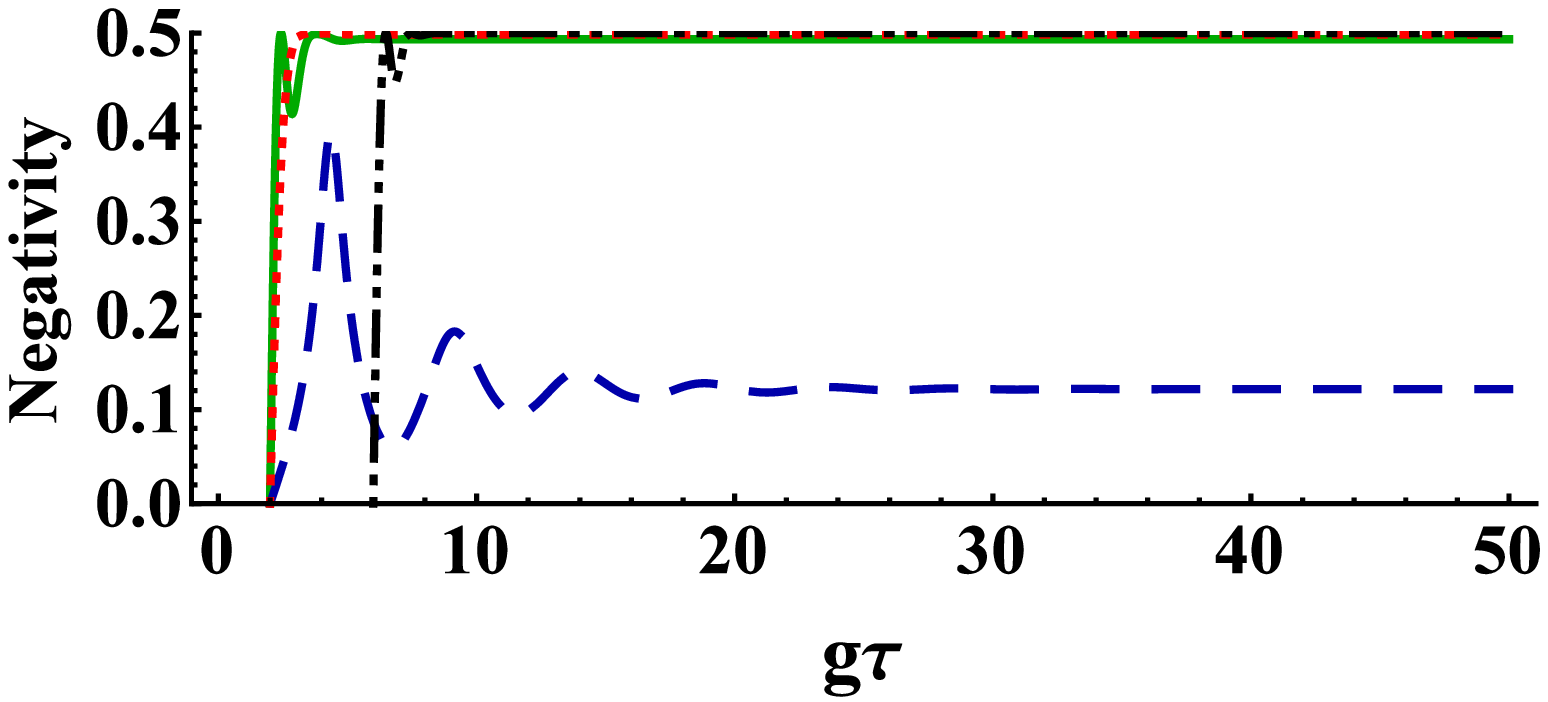}}
    \hspace{0.05\textwidth}
       \subfigure[\label{fig.Fig2e} \ $N'_5(\rho)=N_8(\rho)$]{\includegraphics[width=0.35\textwidth]{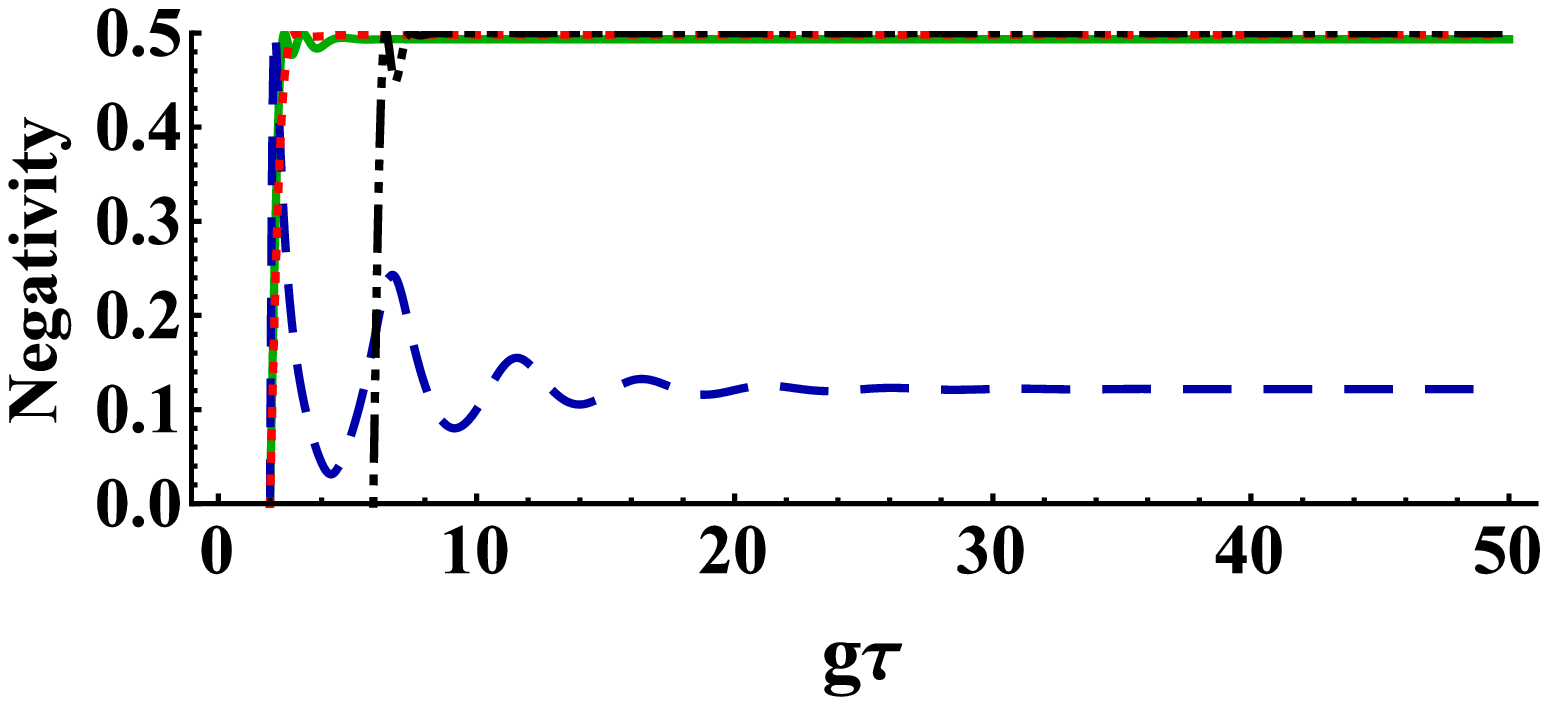}}
        \hspace{0.05\textwidth}
 \subfigure[\label{fig.Fig2f} \ $N_6(\rho)=N'_6(\rho)=N_7(\rho)=N'_7(\rho)$]{\includegraphics[width=0.35\textwidth]{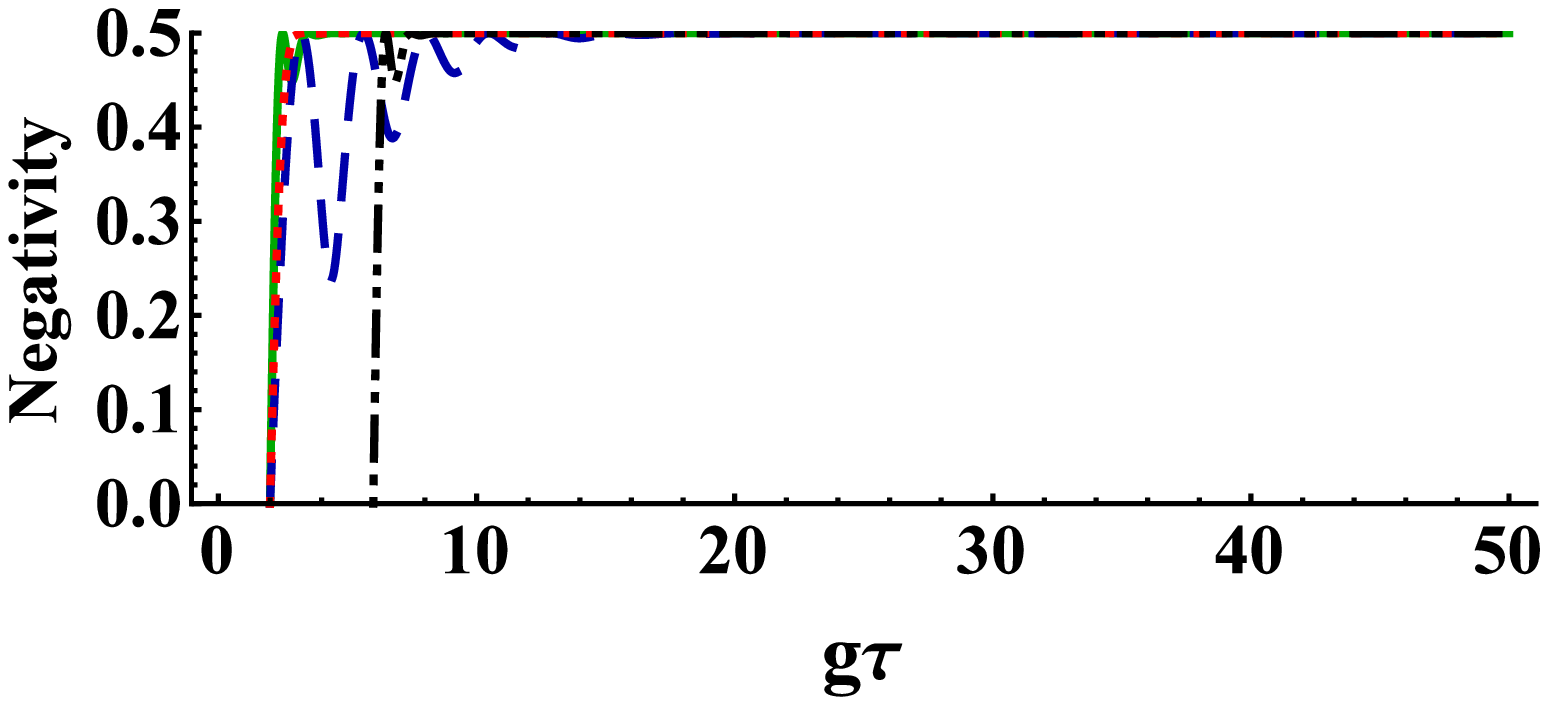}}
 \caption{\label{fig.Negativity} {\it The time evolution of negativity}: (a) $N_1(\rho)=N'_4(\rho)$ (b) $N'_1(\rho)=N_4(\rho)$ (c) $N_2(\rho)=N'_2(\rho)=N_3(\rho)=N'_3(\rho)$ for $\Delta=2g$, $\Gamma=4g$, $gt=2$ (solid green line), for $\Delta=6g$, $\Gamma=4g$, $gt=2$ (dashed blue line), for $\Delta=2g$, $\Gamma=12g$, $gt=2$ (dotted red line), for $\Delta=2g$, $\Gamma=4g$, $gt=6$ (dot-dot-dashed black line) with $g_1=g$ and (d) $N_5(\rho)=N'_8(\rho)$ (e) $N'_5(\rho)=N_8(\rho)$ (f) $N_6(\rho)=N'_6(\rho)=N_7(\rho)=N'_7(\rho)$ for $\delta=2g$, $\gamma=2g$, $gt=2$ (solid green line), for $\delta=6g$, $\gamma=2g$, $gt=2$ (dashed blue line), for $\delta=2g$, $\gamma=6g$, $gt=2$ (dotted red line), for $\delta=2g$, $\gamma=2g$, $gt=6$ (dot-dot-dashed black line) with $g_2=2g$.}
  \end{figure}

\begin{figure}
  \centering
  \subfigure[\label{fig.Fig3a} \ $S_1(\tau)=S'_4(\tau)$]{\includegraphics[width=0.35\textwidth]{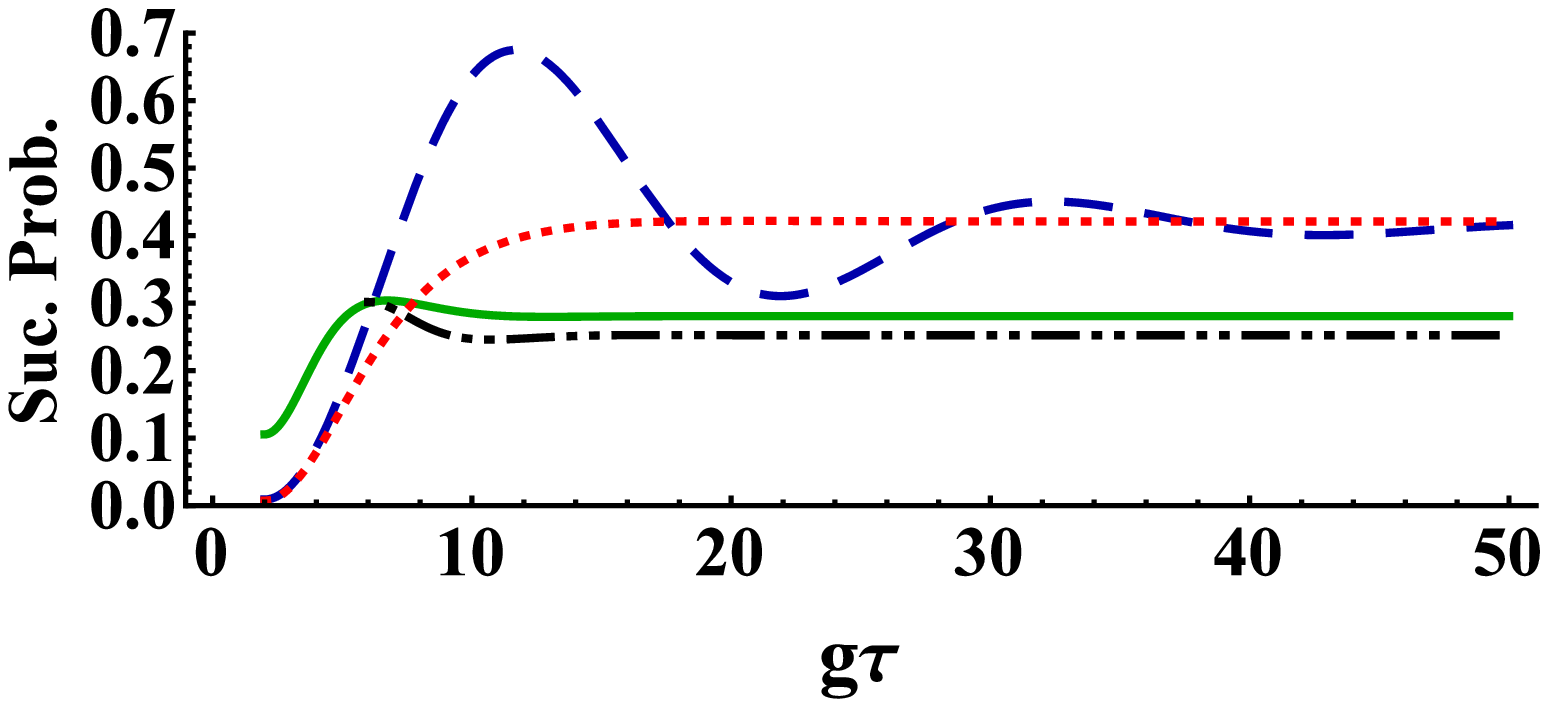}}
  \hspace{0.05\textwidth}
   \subfigure[\label{fig.Fig3b} \ $S'_1(\tau)=S_4(\tau)$]{\includegraphics[width=0.35\textwidth]{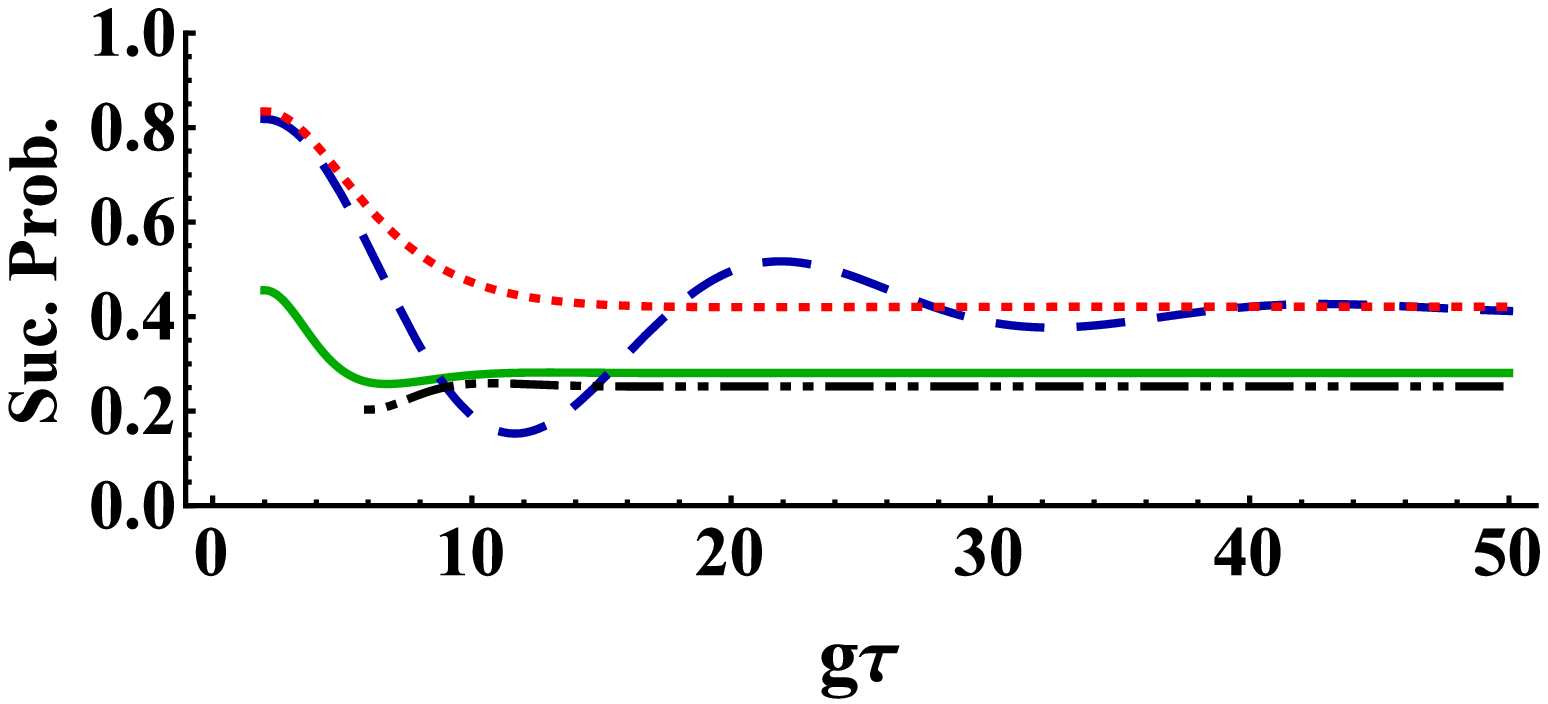}}
   \hspace{0.05\textwidth}
    \subfigure[\label{fig.Fig3c} \ $S_2(\tau)=S'_2(\tau)$]{\includegraphics[width=0.35\textwidth]{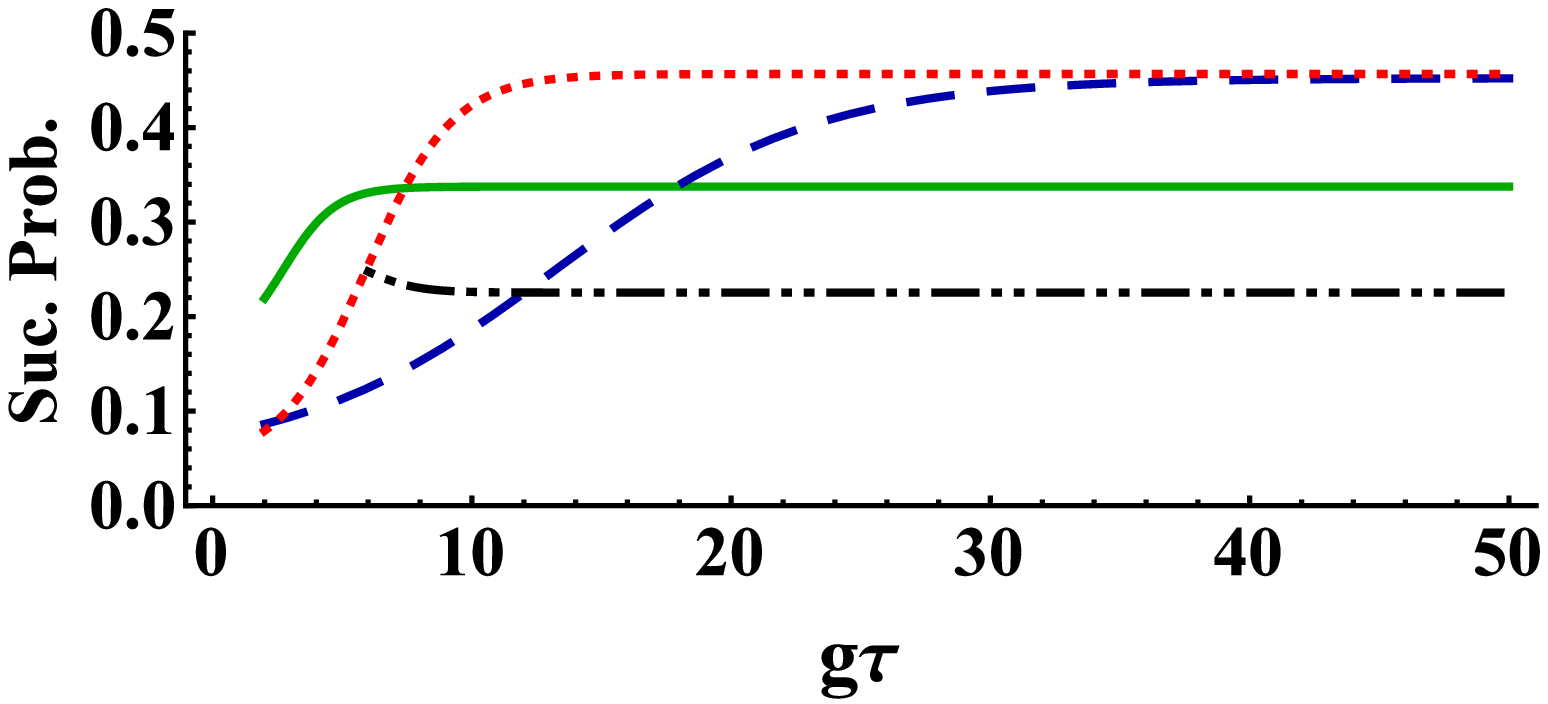}}
    \hspace{0.05\textwidth}
     \subfigure[\label{fig.Fig3d} \ $S_3(\tau)=S'_3(\tau)$]{\includegraphics[width=0.35\textwidth]{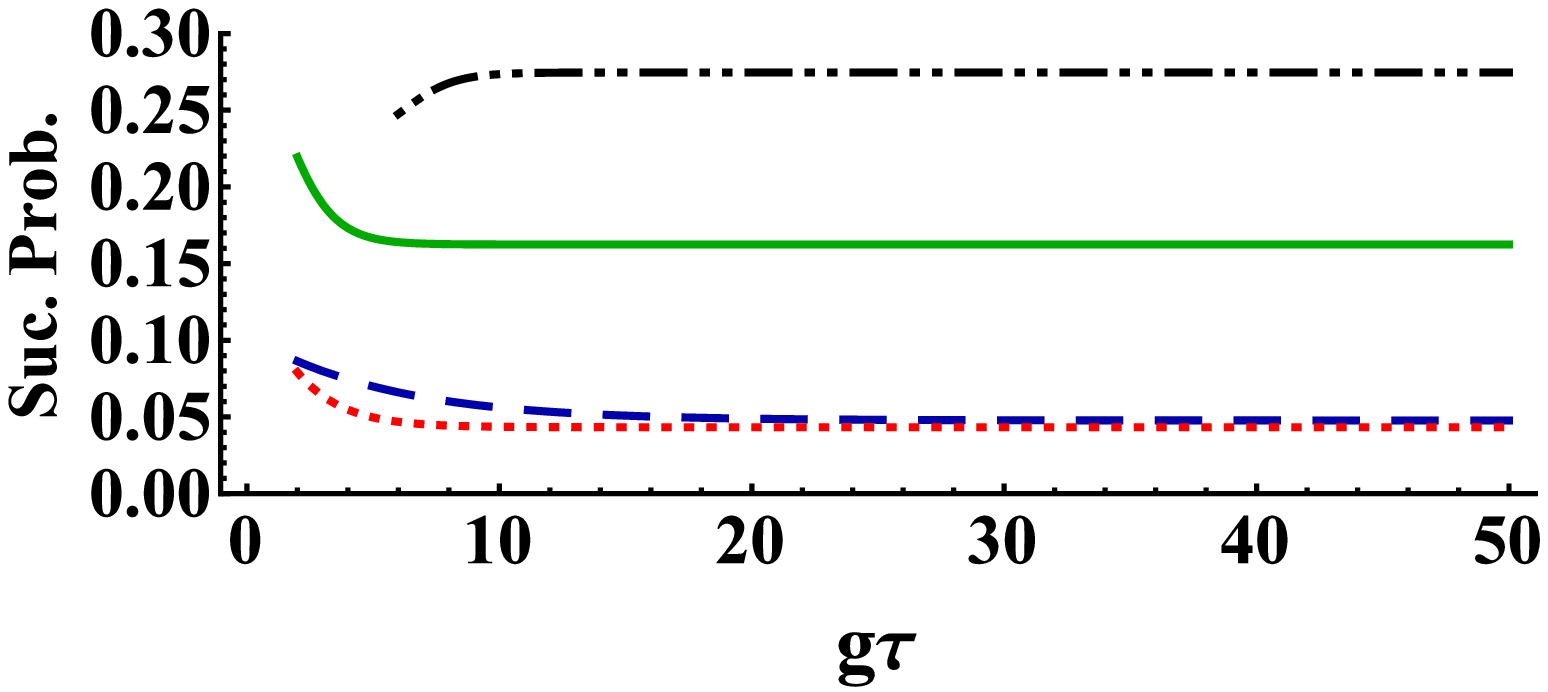}}
        \hspace{0.05\textwidth}
     \subfigure[\label{fig.Fig3e} \ $S_5(\tau)=S'_8(\tau)$]{\includegraphics[width=0.35\textwidth]{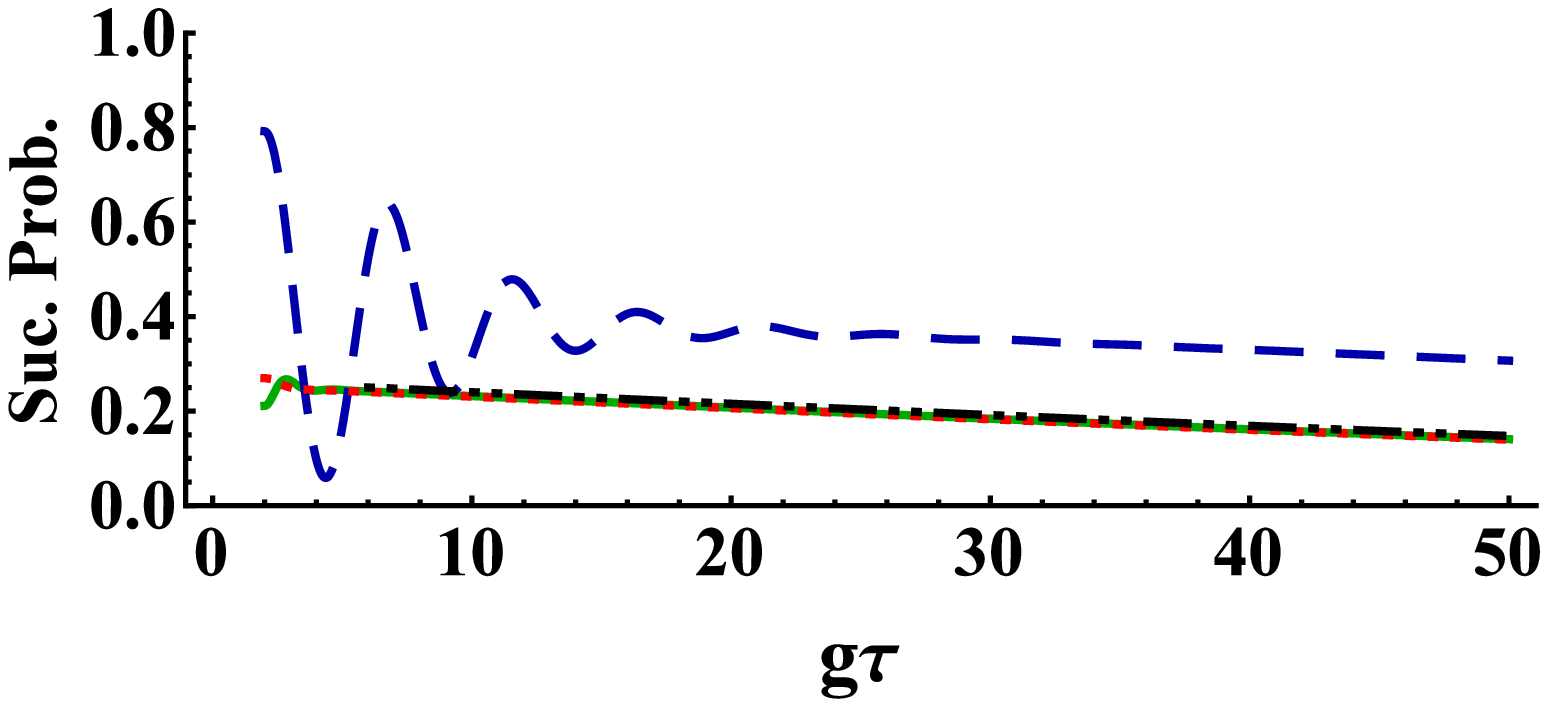}}
     \hspace{0.05\textwidth}
      \subfigure[\label{fig.Fig3f} \ $S'_5(\tau)=S_8(\tau)$]{\includegraphics[width=0.35\textwidth]{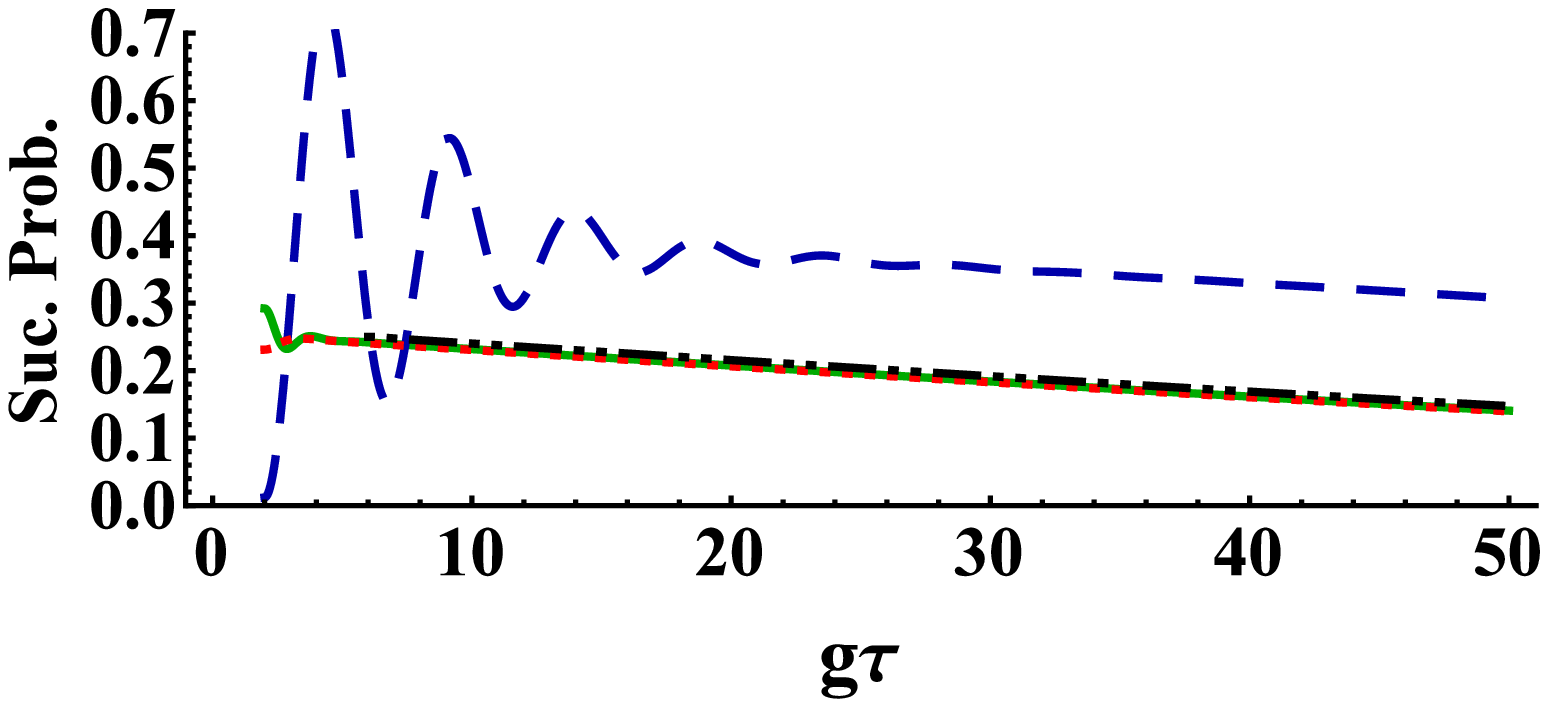}}
          \hspace{0.05\textwidth}
        \subfigure[\label{fig.Fig3g} \ $S_6(\tau)=S'_6(\tau)$]{\includegraphics[width=0.35\textwidth]{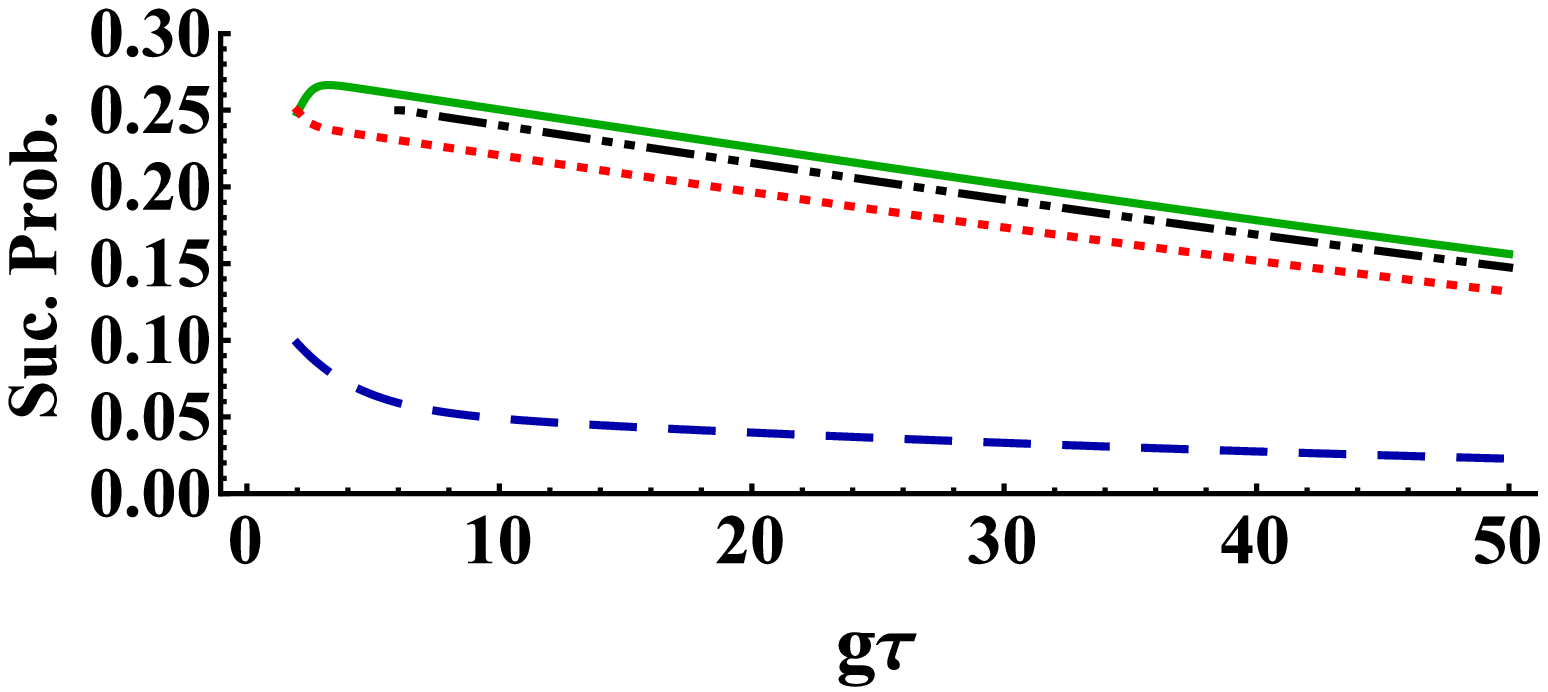}}
         \hspace{0.05\textwidth}
  \subfigure[\label{fig.Fig3h} \ $S_7(\tau)=S'_7(\tau)$]{\includegraphics[width=0.35\textwidth]{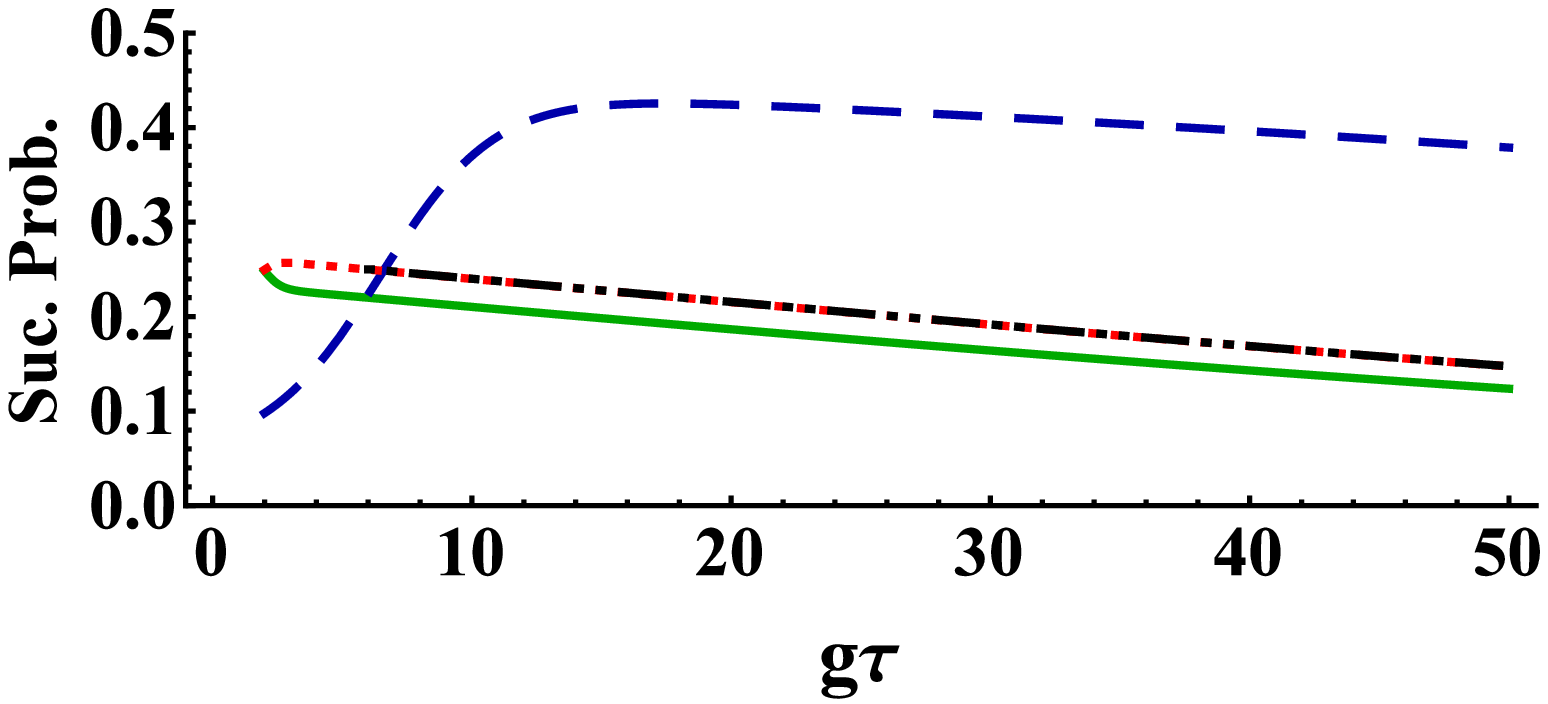}}
  \caption{\label{fig.successprobability} {\it The time evolution of success probability}: (a) $S_1(\tau)=S'_4(\tau)$ (b) $S'_1(\tau)=S_4(\tau)$ (c) $S_2(\tau)=S'_2(\tau)$ (d) $S_3(\tau)=S'_3(\tau)$ for $\Delta=2g$, $\Gamma=4g$, $gt=2$ (solid green line), for $\Delta=6g$, $\Gamma=4g$, $gt=2$ (dashed blue line), for $\Delta=2g$, $\Gamma=12g$, $gt=2$ (dotted red line), for $\Delta=2g$, $\Gamma=4g$, $gt=6$ (dot-dot-dashed black line) with $g_1=g$, $g_2=2g$, $\delta=2g$ and $\gamma=0$ and (e) $S_5(\tau)=S'_8(\tau)$ (f) $S'_5(\tau)=S_8(\tau)$ (g) $S_6(\tau)=S'_6(\tau)$ (h) $S_7(\tau)=S'_7(\tau)$ for $\delta=2g$, $\gamma=2g$, $gt=2$ (solid green line), for $\delta=6g$, $\gamma=2g$, $gt=2$ (dashed blue line), for $\delta=2g$, $\gamma=6g$, $gt=2$ (dotted red line), for $\delta=2g$, $\gamma=2g$, $gt=6$ (dot-dot-dashed black line) with $g_1=g$, $g_2=2g$, $\Delta=10g$ and $\Gamma=2g$.}
 \end{figure}

\begin{figure}
   \centering
   \subfigure[\label{fig.Fig4a} \  $\Delta=6g$, $\Gamma=0$]{\includegraphics[width=0.35\textwidth]{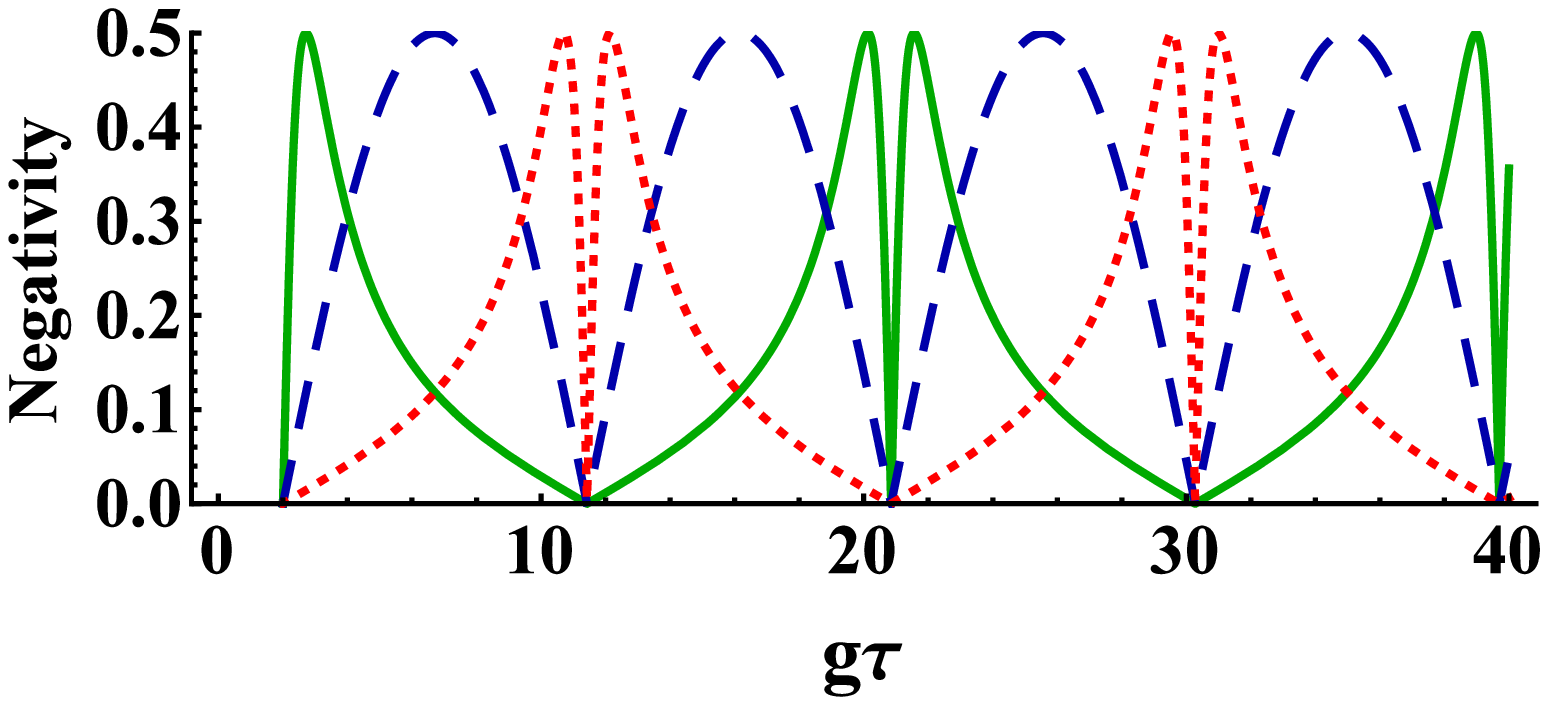}}
   \hspace{0.05\textwidth}
    \subfigure[\label{fig.Fig4b} \ $\delta=6g$, $\gamma=0$]{\includegraphics[width=0.35\textwidth]{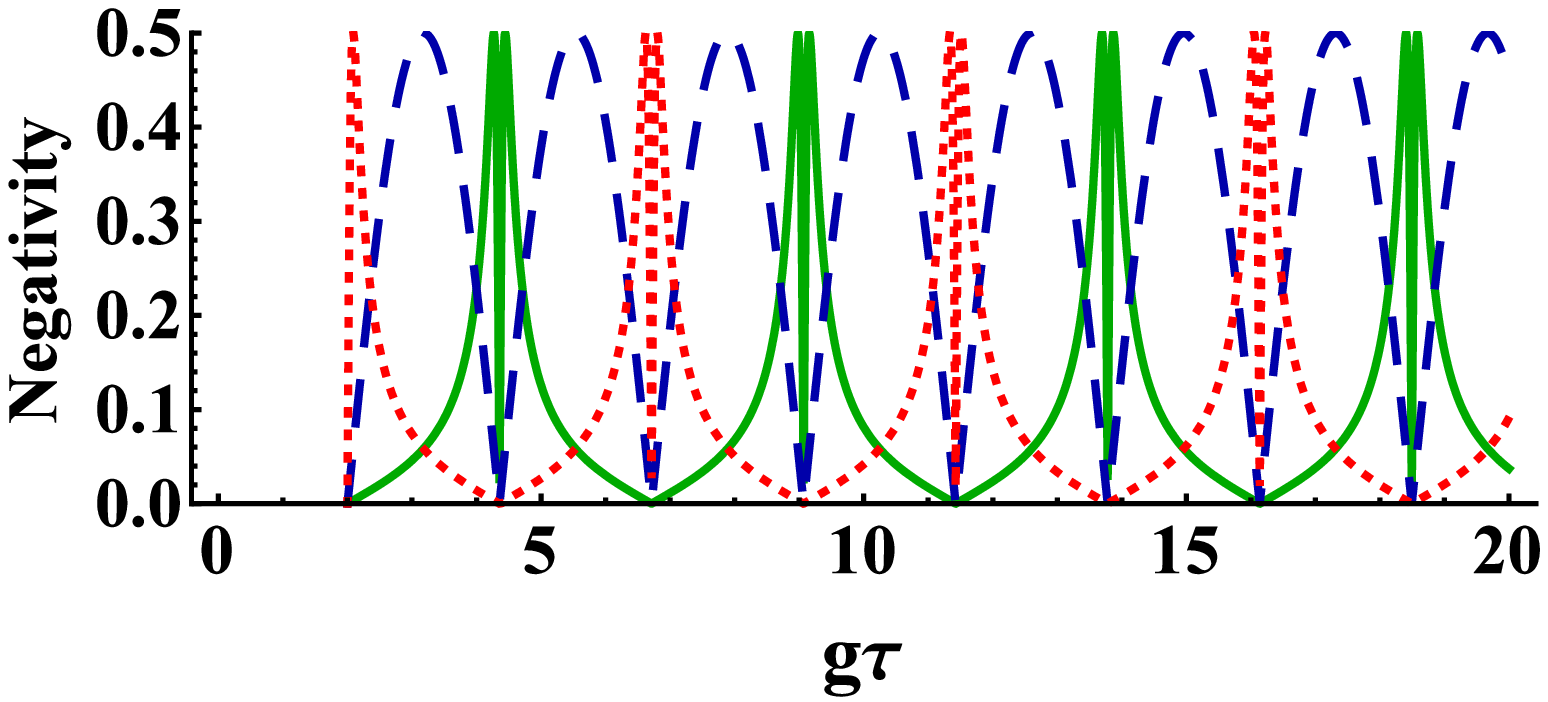}}
     \caption{\label{fig.Negativity2} {\it The time evolution of non-dissipative negativity}: (a) $\Delta=6g$, $\Gamma=0$ for $N_1(\rho)=N'_4(\rho)$ (solid green line), $N_2(\rho)=N'_2(\rho)=N_3(\rho)=N'_3(\rho)$ (dashed blue line) and $N_4(\rho)=N'_1(\rho)$ (dotted red line) (b) $\delta=6g$, $\gamma=0$ for $N_5(\rho)=N'_8(\rho)$ (solid green line), $N_6(\rho)=N'_6(\rho)=N_7(\rho)=N'_7(\rho)$ (dashed blue line) and $N_8(\rho)=N'_5(\rho)$ (dotted red line) with $g_1=g$, $g_2=2g$ and $gt=2$.}
  \end{figure}

Now, which we successfully transferred the entanglement to the distant atoms (1,8), our purpose of this section is to discuss the time evolution of "negativity" and "success probability" of the obtained entangled state of atoms (1,8). In this respect, we analyse the effect of detuning, dissipation and initial interaction time on the above two quantities.\\
  From the closed form of the negativities obtained in previous section, it is found that $N_1(\rho)$, $N'_1(\rho)$, $N_2(\rho)$, $N'_2(\rho)$, $N_3(\rho)$, $N'_3(\rho)$, $N_4(\rho)$ and $N'_4(\rho)$ ($N_5(\rho)$, $N'_5(\rho)$, $N_6(\rho)$, $N'_6(\rho)$, $N_7(\rho)$, $N'_7(\rho)$, $N_8(\rho)$ and $N'_8(\rho)$) are dependent on $\lambda_1$ ($\lambda_2$), but the introduced success probabilities depend on both $\lambda_1$ and $\lambda_2$. So we can achieve suitable success probabilities for each negativities $N_1(\rho)$, $N'_1(\rho)$, $N_2(\rho)$, $N'_2(\rho)$, $N_3(\rho)$, $N'_3(\rho)$, $N_4(\rho)$ and $N'_4(\rho)$ ($N_5(\rho)$, $N'_5(\rho)$, $N_6(\rho)$, $N'_6(\rho)$, $N_7(\rho)$, $N'_7(\rho)$, $N_8(\rho)$ and $N'_8(\rho)$) by justifying $\lambda_2$ ($\lambda_1$).\\
 In figure \ref{fig.Negativity} (\ref{fig.successprobability}) the effect of detuning, dissipation and initial interaction time has been considered on negativity (success probability). In these figures, the dashed blue lines, dotted red lines and dot-dot-dashed black lines that show the effect of detuning, dissipation and initial interaction time, respectively, are compared with solid green lines. Figures \ref{fig.Fig2a}, \ref{fig.Fig2b} and \ref{fig.Fig2c} are independent of $\lambda_2$ (independent of $g_2$, $\gamma$ and $\delta$), and increasing the detuning $\Delta$ and dissipation $\Gamma$ (initial interaction time) leads to decreasing (increasing) the entanglement in figures \ref{fig.Fig2a} and \ref{fig.Fig2b}. On the other hand, as is observed from figure \ref{fig.Fig2c}, the time evolution of negativity has not been effectively changed by increasing detuning $\Delta$, dissipation $\Gamma$ and initial interaction time. Figures \ref{fig.Fig2d}, \ref{fig.Fig2e} and \ref{fig.Fig2f} are independent of $\lambda_1$ (independent of $g_1$, $\Gamma$ and $\Delta$). Increasing the detuning $\delta$ has destructive effect on negativity in these figures, but the other parameters do not critically affect on negativity. Generally, the oscillatory evolution of negativity disappears as time goes on and the entanglement of atoms (1,8) is maintained at some finite values up to its maximum value 0.5.\\
 In figure \ref{fig.successprobability} the time evolution of success probability is considered. In the plots presented like before the dashed blue lines, dotted red lines and dot-dot-dashed black lines show the effect of detuning, dissipation and initial interaction time on success probability, respectively, and these lines are compared with solid green lines. Increasing the detuning $\Delta$ and dissipation  $\Gamma$ (initial interaction time) has constructive (minor destructive) effect on success probability as is shown in figures \ref{fig.Fig3a}, \ref{fig.Fig3b} and \ref{fig.Fig3c}. While in figure \ref{fig.Fig3d} success probability is decreased (increased) by increasing the detuning $\Delta$ and dissipation  $\Gamma$ (initial interaction time). The time evolution of success probability in figures \ref{fig.Fig3e}, \ref{fig.Fig3f}, \ref{fig.Fig3g} and \ref{fig.Fig3h} is 	
 invariant for different chosen parameters of dissipation  $\gamma$ and initial interaction time. In addition, in figures \ref{fig.Fig3e}, \ref{fig.Fig3f} and \ref{fig.Fig3h} the success probability has been increased by increasing the detuning $\delta$.\\
 Finally, the regular periodic evolution of negativity in the absence of dissipation is observable in figures \ref{fig.Fig4a} and \ref{fig.Fig4b}. It is important to notice that the effective atom-field dissipation parameters $\Gamma$ and $\gamma$ are removable if the conditions $\gamma_e-\gamma_g-\kappa=0=\gamma_e-\gamma_f-\kappa'$ are satisfied. In this way we can surprisingly choose the involved parameters such that the atomic and field dissipation effects can be effectively and completely ignored, while they are essentially considered in our model.  Indeed, the involved parameters come to help one to perform a complete ideal quantum repeater while dissipation effects have been considered.

\section{Summary and conclusions} \label{sec.Conclusion}
In this paper the entanglement was transferred over a large distance to two three-level atoms using the quantum repeater protocol. To achieve the purpose, the distance between the two atoms was divided into short parts and insert there six three-level $\Lambda$-type atoms. The pairs (1,2), (3,4), (5,6) and (7,8) have initially been prepared in atomic Bell states. By performing interaction between atoms (2,3) and (6,7) in the dissipative cavities in the presence of atomic spontaneous emission, the entanglement was produced between pairs (1,4) and (5,8). Finally, the entangled pair (1,8) has been achieved by performing dissipative interaction between two three-level atoms (4,5). Also, the effect of detuning, dissipation and initial interaction time was then considered on negativity and success probability. The calculated negativities $N_1(\rho)$, $N'_1(\rho)$, $N_2(\rho)$, $N'_2(\rho)$, $N_3(\rho)$, $N'_3(\rho)$, $N_4(\rho)$ and $N'_4(\rho)$ ($N_5(\rho)$, $N'_5(\rho)$, $N_6(\rho)$, $N'_6(\rho)$, $N_7(\rho)$, $N'_7(\rho)$, $N_8(\rho)$ and $N'_8(\rho)$) are dependent on $\lambda_1$ ($\lambda_2$), but the introduced success probabilities depend on $\lambda_1$ and $\lambda_2$. So the suitable success probabilities for each negativities $N_1(\rho)$, $N'_1(\rho)$, $N_2(\rho)$, $N'_2(\rho)$, $N_3(\rho)$, $N'_3(\rho)$, $N_4(\rho)$ and $N'_4(\rho)$ ($N_5(\rho)$, $N'_5(\rho)$, $N_6(\rho)$, $N'_6(\rho)$, $N_7(\rho)$, $N'_7(\rho)$, $N_8(\rho)$ and $N'_8(\rho)$) can be achieved by controlling $\lambda_2$ ($\lambda_1$). The time evolution of negativity is regularly periodic in the absence of dissipation. The maxima of negativity have been increased by decreasing (increasing) the detuning (dissipation) in most cases. However, increasing the initial interaction time has constructive effect on the negativity in all cases. The oscillation of negativity has been removed as time goes on and the entanglement was then stabled at some finite values in these times. The effective atom-field dissipation parameters $\Gamma$ and $\gamma$  (see Eq.  (\ref{effdiss})) can be completely removed in our proposed model by choosing  atoms and cavities with appropriate spontaneous emission and photon-leakages, respectively.

  \end{document}